\documentclass[preprint,prd,nofootinbib,tightenlines,amsmath]{revtex4}
    
\usepackage{axodraw} 
\usepackage{bm} 
\usepackage{graphicx}
    
\begin{document}

\baselineskip=15pt
\parskip=5pt
        
\vspace*{5ex}
 
\hfill November 2002  
 
\vspace*{5ex}
      
\title{\mbox{$\bm{CP}$}  Violation in Hyperon Nonleptonic Decays \\ within 
the Standard Model}

\author{Jusak Tandean${}^{1,2}$}
\email{jtandean@mail.physics.smu.edu}

\author{G.~Valencia${}^3$} 
\email{valencia@iastate.edu} 

\affiliation{
${}^1$Department of Physics and Astronomy, University of Kentucky,\\
Lexington, Kentucky 40506-0055
}

\affiliation{  
${}^2$Department of Physics, Southern Methodist University,   
Dallas, Texas 75275-0175\footnote{Present address.}  
\vspace{2ex} \\} 

\affiliation{  
${}^3$Department of Physics and Astronomy, Iowa State University, \\ 
Ames, Iowa 50011
\vspace{5ex} \\}
    

\begin{abstract}  
We calculate the $CP$-violating asymmetries $A(\Lambda^0_-)$ 
and $A(\Xi^-_-)$ in nonleptonic hyperon decay within the Standard Model 
using the framework of heavy-baryon chiral perturbation 
theory ($\chi$PT). We identify those terms that correspond to previous 
calculations and discover several errors in the existing 
literature. We present a new result for the lowest-order 
(in $\chi$PT) contribution of the penguin operator to these 
asymmetries, as well as an estimate for the uncertainty of our 
result that is based on the calculation of the leading 
nonanalytic corrections.     
\end{abstract}   
     

\maketitle

\section{Introduction}

In nonleptonic hyperon decays such as  $\,\Lambda\to p\pi^-,\,$ 
it is possible to search for $CP$ violation by comparing the 
angular distribution with that of the corresponding 
anti-hyperon decay~\cite{op}. 
The Fermilab experiment HyperCP is currently analyzing data 
searching for $CP$ violation in such a decay.
   
The reaction of interest for HyperCP is the decay of a polarized 
$\Lambda$, with known polarization  $\bm{w}$,  
into a proton (whose polarization is not measured) and 
a $\pi^-$ with momentum  $\bm{q}$. 
The interesting observable is a correlation in the decay 
distribution of the form
\begin{equation}
\frac{{\rm d}\Gamma}{{\rm d}\Omega}  \,\sim\,  
1 +\alpha\, \bm{w}\cdot\bm{q}   \,\,.  
\end{equation}
The branching ratio for this mode is $63.9\%$,  and 
the parameter $\alpha$ has been measured to be  
$\,\alpha_\Lambda^{} = 0.642\,$~\cite{pdb}.
The $CP$ violation in question involves a comparison of 
the parameter $\alpha$ with the corresponding parameter 
$\bar{\alpha}$ from the reaction  
$\,\bar{\Lambda}\to \bar{p}\pi^+.\,$

To obtain polarized $\Lambda$'s with known polarization, it is 
necessary to study the double decay chain  
$\,\Xi^- \to \Lambda \pi^-\to p \pi^- \pi^-\,$~\cite{e756,hypercp}.  
This eventually leads to the experimental observable being 
sensitive to the {\it sum} of $CP$ violation in 
the $\Xi$ decay and $CP$ violation in the $\Lambda$ decay.

In both reactions,  $\,\Xi^-\to \Lambda \pi^-\,$  and  
$\,\Lambda\to p \pi^-,\,$  the final state can be reached from  
the initial state via  $\,|\Delta I|=\frac{1}{2}\,$  or  
$\,|\Delta I|=\frac{3}{2}\,$  transitions. 
It is known that due to the existence of a strong 
$\,|\Delta I|=\frac{1}{2}\,$  rule for nonleptonic hyperon decay, 
the dominant contribution to the  $CP$-violating asymmetries 
arises from interference between  an $S$-wave and  a $P$-wave 
within the  $\,|\Delta I|=\frac{1}{2}\,$  
transition~\cite{dp,dhp,hsv}. 
One can define the $CP$-violating asymmetries  
\begin{equation}
A_\Lambda^{}  \,\equiv\,   
A\bigl(\Lambda^0_-\bigr)  \,\equiv\,  
\frac{\alpha_{\Lambda}^{}-\bar{\alpha}_{\Lambda}^{}}{
\alpha_{\Lambda}^{}+\bar{\alpha}_{\Lambda}^{} }   \,\,,   
\hspace{3em}  
A_\Xi^{}  \,\equiv\,  A\bigl(\Xi^-_-\bigr)  \,\equiv\,  
\frac{\alpha_\Xi^{}-\bar{\alpha}_\Xi^{}}{
\alpha_\Xi^{}+\bar{\alpha}_\Xi^{} }   
\end{equation}  
for the $\Lambda$  and  $\Xi^-$  decays,  respectively.   
The experimental observable is then~\cite{e756,hypercp}   
\begin{equation}
A_{\Xi\Lambda}^{}  \,\simeq\,  A_\Lambda^{} + A_\Xi^{}  \,\,.
\label{adefin}
\end{equation} 
Approximate expressions have been obtained for $A_{\Lambda,\Xi}^{}$ 
in the case of  $\,|\Delta I|=\frac{1}{2}\,$  dominance~\cite{dhp}, namely
\begin{equation}   \label{approxa}
\begin{array}{c}   \displaystyle
A_{\Lambda}^{}  \,\simeq\,    
- \tan \left(\delta_P^\Lambda -\delta_S^\Lambda\right) 
\sin\left(\phi_P^\Lambda-\phi_S^\Lambda\right)  \,\,,  
\vspace{2ex} \\   \displaystyle
A_{\Xi}^{}  \,\simeq\,    
- \tan \left(\delta_P^\Xi -\delta_S^\Xi\right) 
\sin\left(\phi_P^\Xi-\phi_S^\Xi\right)  \,\,.  
\end{array}
\end{equation}
Here, $\delta_{S}^\Lambda$ $\bigl(\delta_{P}^\Lambda\bigr)$  is the 
strong $S$-wave ($P$-wave)  $N\pi$ scattering phase-shift at  
$\,\sqrt s=M_\Lambda^{},\,$  and  
$\delta_{S}^\Xi$ $\bigl(\delta_{P}^\Xi\bigr)$ 
is the strong $S$-wave ($P$-wave) $\Lambda \pi$ scattering 
phase-shift at $\,\sqrt s=M_\Xi^{}.\,$ 
Moreover,  $\phi_{S}^{\Lambda,\Xi}$  $\bigl(\phi_{P}^{\Lambda,\Xi}\bigr)$  
are the $CP$-violating weak phases induced by the $\,|\Delta S|=1,\,$  
$|\Delta I|=\frac{1}{2}\,$  interaction in the $S$-wave ($P$-wave) of 
the  $\,\Lambda\to p\pi^-\,$ and  $\,\Xi^-\to\Lambda\pi^-\,$  decays, 
respectively.

Experimentally, the current published limit is 
$\,A_{\Xi\Lambda}^{}=0.012\pm 0.014\,$  from E756~\cite{e756}, 
and the expected sensitivity of HyperCP is  $10^{-4}\,$~\cite{hypercp}.   
In addition, HyperCP has recently obtained a preliminary measurement 
of  $\,A_{\Xi\Lambda}^{}=(-7\pm12\pm6.2)\times 10^{-4}\,$~\cite{zyla}.
Previous estimates for $A_{\Xi\Lambda}^{}$ indicated that it 
occurs at the few times $10^{-5}$ level within the standard 
model~\cite{dhp,hsv,im}  and that it can be as large as $10^{-3}$ 
beyond the standard model~\cite{dhp,NP}. 
The larger asymmetries occur in models with an enhanced gluon-dipole  
operator that is parity-even and thus does not contribute to the 
$\epsilon^\prime$  parameter in kaon decay.  
The $10^{-3}$ upper bound corresponds to the phenomenological 
constraint from new contributions to the $\epsilon$ parameter in 
kaon mixing.  
This illustrates the relevance of the HyperCP  measurement which  
complements the $\epsilon^\prime$ experiments in the study of $CP$ 
violation in  $\,|\Delta S|=1\,$  transitions.

The strong  $\pi N$  scattering phases needed in Eq.~(\ref{approxa}) 
have been measured to be  $\,\delta_S^\Lambda\sim 6^\circ\,$  
and  $\,\delta_P^\Lambda\sim -1^\circ\,$  with errors of 
about~$1^\circ\,$~\cite{roper}. 
In contrast, the strong  $\Lambda \pi$ scattering phases have not 
been measured.   
Modern calculations based on chiral perturbation theory indicate 
that these phases are small,  with  $\bigl|\delta_S^\Xi\bigr|$ 
being at most~$7^\circ\,$~\cite{lsw,DatPak,kamal,ttv,mo,kaiser}. 
For our numerical estimates, we will allow the  $\Lambda \pi$ 
phases to vary within the range obtained at next-to-leading order  
in heavy-baryon chiral perturbation theory~\cite{ttv},
\begin{eqnarray}   
\begin{array}{c}   \displaystyle   
-3.0^\circ  \,\le\,  \delta_S^\Xi  \,\le\,  +0.4^\circ   \,\,,    
\hspace{3em}   
-3.5^\circ  \,\le\,  \delta_P^\Xi  \,\le\,  -1.2^\circ   \,\,.  
\end{array}     
\end{eqnarray}     
One could choose to be less constrained and include the larger  
$\,\delta_S^\Xi=-7^\circ\,$  found in Ref.~\cite{ttv}, but this 
would only enlarge the  $\delta_S^\Xi$ range and hence  
the uncertainty of the predicted asymmetry.    
In any case, eventually these phases can be extracted directly  
from the measurement of the decay distribution in  
$\,\Xi\to\Lambda \pi\,$~\cite{hypercp,alak,huang}.  
Recently E756 has reported a preliminary result of  
$\,\delta_P^\Xi-\delta_S^\Xi=3.17^\circ\pm5.45^\circ\,$~\cite{alak}.

In this paper, we estimate the weak phases that appear in  
$A_\Lambda^{}$  and  $A_\Xi^{}$  within the standard model. 
In Sec.~\ref{cpt}, we present a calculation of the weak phases 
guided by heavy-baryon chiral perturbation theory in terms of 
three unknown weak counterterms. 
In Sec.~\ref{ect}, we estimate the value of these 
counterterms by considering contributions that arise from the 
factorization of the penguin operator and also nonfactorizable 
contributions estimated in the MIT bag model. 
Sec.~\ref{results} contains the resulting weak phases and 
$CP$-violating asymmetries.  
Finally, in Sec.~\ref{disc}, we compare our results to those 
of previous work and present our conclusions.
For completeness, we also provide in an appendix the results for 
the corresponding asymmetries in $\,\Sigma\to N\pi\,$  decays.

\section{Chiral perturbation theory\label{cpt}}
     
The chiral Lagrangian that describes the interactions of 
the lowest-lying mesons and baryons is written down in terms of 
the lightest meson-octet, baryon-octet, and baryon-decuplet 
fields~\cite{bsw,georgi,dgh,JenMan1}.  
The meson and baryon octets are collected into  $3\times3$  matrices   
$\varphi$  and~$B$,  respectively, and the decuplet fields are 
represented by the Rarita-Schwinger  tensor~$T_{abc}^\mu$,  
which is completely symmetric in its SU(3) indices ($a,b,c$).  
The octet mesons enter through the exponential  
$\,\Sigma=\xi^2=\exp({\rm i}\varphi/f),\,$  where  $f$  
is the pion-decay constant.

In the heavy-baryon formalism~\cite{JenMan1,JenMan2}, the baryons 
in the chiral Lagrangian are described by velocity-dependent 
fields,  $B_v^{}$  and  $T_v^\mu$.   
For the strong interactions, the leading-order Lagrangian 
is given  by~\cite{JenMan1,JenMan2,JenMan3}
\begin{eqnarray}   \label{Ls1}   
{\cal L}_{\rm s}^{(1)}  &=&  
\mbox{$\frac{1}{4}$} f^2 \left\langle  
\partial^\mu\Sigma^\dagger\, \partial_\mu^{}\Sigma \right\rangle     
+ \left\langle \bar B_v^{}\, {\rm i}v\cdot{\cal D} B_v^{} 
 \right\rangle     
+ 2D \left\langle \bar B_v^{} S_v^\mu 
 \left\{ {\cal A}_\mu^{}, B_v^{} \right\} \right\rangle 
+ 2F \left\langle \bar B_v^{} S_v^\mu  
 \left[ {\cal A}_\mu^{}, B_v^{} \right] \right\rangle    
\nonumber \\ &&   
-\,\,    
\bar T_v^\mu {\rm i} v\cdot {\cal D} T_{v\mu}^{}  
+ \Delta m\, \bar T_v^\mu T_{v\mu}^{}  
+ {\cal C} \left( \bar T_v^\mu {\cal A}_\mu^{} B_v^{} 
                   + \bar B_v^{} {\cal A}_\mu^{} T_v^\mu \right)    
+ 2{\cal H}\, \bar T_v^\mu S_v^{}\cdot{\cal A} T_{v\mu}^{}   \,\,,       
\end{eqnarray}      
where  $\langle\cdots\rangle$ denotes  ${\rm Tr}(\cdots)$  
in flavor-SU(3) space,  $S_v^{}$  is the spin operator, and   
\begin{eqnarray}   
\begin{array}{c}   \displaystyle  
{\cal A}_\mu^{}  \,=\,  
\mbox{$\frac{\rm i}{2}$}  
\left( \xi\, \partial_\mu^{}\xi^\dagger 
      - \xi^\dagger\, \partial_\mu^{}\xi \right) 
\,=\,  
\frac{\partial_\mu^{}\varphi}{2f}  \,\,+\,\,  {\cal O}(\varphi^3)  \,\,,  
\end{array}   
\end{eqnarray}    
with further details given in Ref.~\cite{atv}.  
In this Lagrangian, $D$, $F$, $\cal C$, and $\cal H$  are 
free parameters, which can be determined from hyperon semileptonic 
decays and from strong decays of the form  $\,T\to B\phi.\,$  
Fitting tree-level formulas, one extracts~\cite{JenMan1,JenMan2}
\begin{eqnarray}   
\begin{array}{c}   \displaystyle   
D  \,=\,  0.80  \,\,,   \hspace{3em}     
F  \,=\,  0.50  \,\,,   \hspace{3em}     
|{\cal C}|  \,=\,  1.7  \,\,,  
\end{array}     
\label{sptree}
\end{eqnarray}     
whereas  $\cal H$  is undetermined from this fit.
From the nonrelativistic quark models, one finds the 
relations~\cite{JenMan3} 
\begin{eqnarray}   
\begin{array}{c}   \displaystyle   
3 F  \,=\,  2 D  \,\,,   \hspace{2em}  
{\cal C}  \,=\,  -2D  \,\,,  \hspace{2em}  
{\cal H}  \,=\,  -3D  \,\,,     
\end{array}     
\end{eqnarray}     
which are well satisfied by  $D$, $F$, and  $\cal C$,  
suggesting the tree-level value  
\begin{eqnarray}   \label{Htree}
{\cal H}  \,=\,  -2.4   \,\,.   
\end{eqnarray}     
In our numerical estimates, we use Eqs.~(\ref{sptree}) 
and~(\ref{Htree})  for the leading-order results and 
the estimate of their uncertainty from one-loop contributions, 
with $\cal C$ and $\cal H$ only appearing in loop diagrams 
involving decuplet baryons.   
As another estimate of the uncertainty in these results, we 
will evaluate the effect of varying $D$ and $F$ between 
their tree-level values above and their one-loop values to be 
given later.

At next-to-leading order, the strong Lagrangian contains 
a greater number of terms~\cite{L2refs}.  
The ones of interest here are those that explicitly break 
chiral symmetry, containing one power of the quark-mass matrix  
$\,M={\rm diag}\bigl(0,0,m_s^{}\bigr).\,$  
For our calculation of the factorization of the penguin operator,  
we will need these terms in the form  
\begin{eqnarray}   \label{Ls2}   
{\cal L}_{\rm s}^{(2)}  &=&    
\mbox{$\frac{1}{4}$} f^2 \left\langle \chi_+^{} \right\rangle     
+ \frac{b_D^{}}{2 B_0^{}} \left\langle \bar B_v^{}   
\left\{ \chi_+^{}, B_v^{} \right\} \right\rangle   
+ \frac{b_F^{}}{2 B_0^{}} \left\langle \bar B_v^{}   
 \left[ \chi_+^{}, B_v^{} \right] \right\rangle     
+ \frac{b_0^{}}{2 B_0^{}} \left\langle \chi_+^{} \right\rangle
 \left\langle \bar B_v^{} B_v^{} \right\rangle     
\nonumber \\ &&  
+\,\,  
\frac{c}{2 B_0^{}}\, \bar T_v^\mu \chi_+^{} T_{v\mu}^{}   
- \frac{c_0^{}}{2 B_0^{}} \left\langle \chi_+^{} \right\rangle  
 \bar T_v^\mu T_{v\mu}^{}   \,\,,  
\end{eqnarray}  
where we have used the notation  
$\,\chi_+^{}=\xi^\dagger\chi\xi^\dagger+\xi\chi^\dagger\xi\,$  
to introduce coupling to external (pseudo)scalar sources,  
$\,\chi= s+ip,\,$ such that in the absence of the external 
sources  $\chi$ reduces to the mass matrix,  $\,\chi=2B_0^{} M.\,$  
As will be discussed in the next section, we also need 
from the meson sector the next-to-leading-order Lagrangian    
\begin{eqnarray}   \label{Ls4}
{\cal L}_{\rm s}^{(4)}  \,=\,
L_{5}^{} \left\langle \partial^\mu\Sigma^\dagger\, 
\partial_\mu^{}\Sigma\, \xi^\dagger \chi_+^{} \xi \right\rangle    
\,\,+\,\,  \cdots   \,\,,   
\end{eqnarray}
where only the relevant term is explicitly shown.     
In Eqs.~(\ref{Ls2}) and~(\ref{Ls4}), the constants  $B_0^{}$, 
$b_{D,F,0}^{}$, $c$, $c_0^{}$, and  $L_5^{}$  are free parameters 
to be fixed from data.

As is well known, the weak interactions responsible for hyperon 
nonleptonic decays are described by a  $\,|\Delta S|=1\,$   
Hamiltonian that transforms as  
$\bigl(8_{\rm L}^{},1_{\rm R}^{}\bigr)\oplus
\bigl(27_{\rm L}^{},1_{\rm R}^{}\bigr)$  
under  SU(3$)_{\rm L}^{}$$\times$SU(3$)_{\rm R}^{}$  rotations. 
It is also known from  experiment that the octet term dominates 
the 27-plet term, as indicated by the fact that the  
$\,|\Delta I|=\frac{1}{2}\,$  components of the decay amplitudes are 
larger than the  $\,|\Delta I|=\frac{3}{2}\,$  components by about 
twenty times~\cite{atv,over}.  
We shall, therefore, assume in what follows  that the decays are 
completely characterized by the  $(8_{\rm L}^{},1_{\rm R}^{})$,  
$\,|\Delta I|=\frac{1}{2}\,$  interactions.  
The leading-order chiral Lagrangian for such interactions  
is~\cite{bsw,jenkins2}     
\begin{eqnarray}  
{\cal L}_{\rm w}^{}  &=&     
h_D^{} \left\langle \bar B_v^{} \left\{ 
\xi^\dagger h \xi\,,\,B_v^{} \right\} \right\rangle   
+ h_F^{} \left\langle \bar B_v^{} \left[ 
\xi^\dagger h \xi\,,\,B_v^{} \right] \right\rangle   
+ h_C^{}\, \bar T_v^\mu\, \xi^\dagger h \xi\, T_{v\mu}^{} 
\nonumber \\ &&  
+\,\, 
\gamma_8^{} f^2 \left\langle h\, \partial_\mu^{} \Sigma\,  
\partial^\mu \Sigma^\dagger \right\rangle   
\,\,+\,\,  {\rm H.c.}   \,\,,
\label{weakcl} 
\end{eqnarray}  
where  $h$  is a  3$\times$3  matrix with elements  
$\,h_{ij}^{}=\delta_{i2}^{}\delta_{3j}^{},\,$  and  the parameters 
$h_{D,F,C}^{}$  and  $\gamma_8^{}$  contain the weak phases 
to be discussed below.

The weak Lagrangian in Eq.~(\ref{weakcl}) is thus the leading-order 
(in $\chi$PT) realization of the effective  $\,|\Delta S|=1\,$  
Hamiltonian in the standard model~\cite{buras}, 
\begin{eqnarray}   \label{weaksd}  
{\cal H}_{\rm w}^{}  \,=\,  
\frac{G_{\rm F}^{}}{\sqrt 2}\, V_{ud}^* V_{us}^{}\,  
\sum_{i=1}^{10} C_i^{}\, Q_i^{}   
\,\,+\,\,  {\rm H.c}   \,\,,    
\end{eqnarray}      
where   $G_{\rm F}^{}$ is the Fermi coupling constant,  
$V_{kl}^{}$ are the elements of the Cabibbo-Kobayashi-Maskawa 
(CKM) matrix~\cite{ckm}, 
\begin{eqnarray}   
C_i^{}  \,\equiv\,  z_i^{} + \tau y_i^{}  \,\equiv\,  
z_i^{}  \,-\,  
\frac{V_{td}^* V_{ts}^{}}{V_{ud}^* V_{us}^{}}\,\, y_i^{}    
\end{eqnarray}      
are the Wilson coefficients,  and  $Q_i^{}$  
are four-quark operators whose expressions can be found 
in Ref.~\cite{buras}.    
Later on, we will express $V_{kl}^{}$  in the Wolfenstein 
parametrization~\cite{wolfenstein}.  
It follows that  
\begin{eqnarray}   
V_{ud}^* V_{us}^{}  \,=\,  \lambda   \,\,,  
\hspace{3em}   
V_{td}^* V_{ts}^{}  \,=\,  -\lambda^5 A^2\, (1-\rho+{\rm i}\eta)  
\end{eqnarray}      
at lowest order in  $\lambda$.       
For our numerical estimates, the relevant parameters that we 
will employ are~\cite{ckmfit}  
\begin{eqnarray}   \label{ckm}
\lambda  \,=\,  0.2219  \,\,,  \hspace{2em}  
A  \,=\,  0.832   \,\,,  \hspace{2em}  
\eta  \,=\,  0.339   \,\,.   
\end{eqnarray}      
In the next section, we match the penguin operator $Q_6^{}$ in the 
short-distance Hamiltonian of Eq.~(\ref{weaksd}) with the 
corresponding Lagrangian parameters in Eq.~(\ref{weakcl}).

\begin{figure}[t]         
\includegraphics{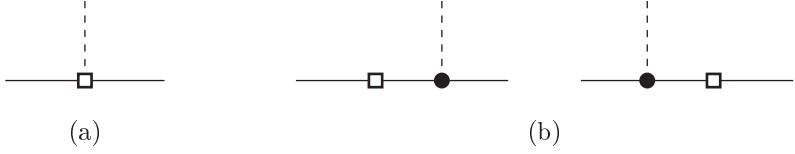}   \vspace{-3ex}
\caption{\label{tree0}%
Leading-order diagrams for (a) $S$-wave and (b) $P$-wave hyperon 
nonleptonic decays. 
In all figures, a solid (dashed) line denotes a baryon-octet 
(meson-octet) field, and a solid dot (hollow square) represents 
a strong (weak) vertex, with the strong vertices being generated 
by  ${\cal L}_{\rm s}^{(1)}$  in Eq.~(\ref{Ls1}).  
Here the weak vertices come from the  $h_{D,F}^{}$  terms in  
Eq.~(\ref{weakcl}).}
\end{figure}

We now have all the ingredients necessary to calculate the weak decay 
amplitudes in terms of the four parameters $h_{D,F,C}^{}$  and
$\gamma_8^{}$ (only the first two are needed at leading order). 
In the heavy-baryon formalism, the amplitude for the weak decay 
of a spin-$\frac{1}{2}$  baryon  $B$  into another   
spin-$\frac{1}{2}$  baryon $B'$  and  a pseudoscalar meson  
$\phi$  has the general form~\cite{jenkins2}    
\begin{eqnarray}   \label{M,HB}     
{\rm i} {\cal M}_{B\to B'\phi}^{}  \,=\,    
-{\rm i} \bigl\langle B'\phi \bigr| {\cal L}_{\rm w+s}^{} 
\bigl| B \bigr\rangle    
\,=\,    
\bar u_{B'}^{}\, \Bigl( {\cal A}_{BB'\phi}^{(S)} 
+ 2S_v^{}\!\cdot\!p_\phi^{}\, {\cal A}_{BB'\phi}^{(P)} 
\Bigr) \, u_{B}^{}   \,\,,  
\end{eqnarray}    
where the superscripts refer to the  $S$- and $P$-wave components 
of the amplitude.  
To express our results, we also adopt the notation~\cite{jenkins2}      
\begin{eqnarray}        
{a}^{(S,P)}_{BB'\phi}  \,\equiv\,  
\sqrt 2\, f^{}\, {\cal A}^{(S,P)}_{BB'\phi}   \,\,.
\label{conamp}   
\end{eqnarray}    
With the Lagrangians given above, one can derive the amplitudes 
at leading order, represented by the diagrams in Fig.~\ref{tree0}. 
Fig.~\ref{tree0}(a) indicates that the $S$-wave is directly 
obtained from a weak vertex provided by Eq.~(\ref{weakcl}). 
The leading contribution to the $P$-wave arises from baryon-pole 
diagrams, as in Fig.~\ref{tree0}(b), which each involve 
a weak vertex from Eq.~(\ref{weakcl}) and a strong vertex from 
Eq.~(\ref{Ls1}). 
Thus the leading-order results for amplitudes not related by 
isospin are~\cite{bsw,jenkins2}   
\begin{subequations}        
\begin{eqnarray}   \label{stree}     
\begin{array}{c}   \displaystyle  
{a}^{(S)}_{\Lambda p\pi^-}  \,=\,  
\mbox{$\frac{1}{\sqrt{6}}$} \bigl( h_D^{}+3 h_F^{} \bigr)    \,\,,   
\hspace{2em}         
{a}^{(S)}_{\Xi^-\Lambda\pi^-}  \,=\,  
\mbox{$\frac{1}{\sqrt{6}}$} \bigl( h_D^{}-3 h_F^{} \bigr)    \,\,,           
\vspace{2ex} \\   \displaystyle    
{a}^{(S)}_{\Sigma^+ n\pi^+}  =  0   \,\,,   
\hspace{2em}      
{a}^{(S)}_{\Sigma^- n\pi^-}  =  -h_D^{} + h_F^{}   \,\,, 
\end{array}    
\end{eqnarray}    
\begin{eqnarray}        
\begin{array}{c}   \displaystyle  
{a}^{(P)}_{\Lambda p\pi^-}     \,=\,   \displaystyle      
\frac{ 2 D\, \bigl( h_D^{}-h_F^{} \bigr)  }{
\sqrt{6}\,\, \bigl( m_\Sigma^{}-m_N^{} \bigr) }   
+ \frac{ (D+F)\, \bigl( h_D^{}+3 h_F^{} \bigr)  
   }{  \sqrt{6}\,\, \bigl( m_\Lambda^{}-m_N^{} \bigr) }   \,\,, 
\vspace{2ex} \\   \displaystyle    
{a}^{(P)}_{\Xi^-\Lambda\pi^-}  \,=\,   \displaystyle   
\frac{ -2 D\, \bigl( h_D^{}+h_F^{} \bigr)  
}{  \sqrt{6}\,\, \bigl( m_\Xi^{}-m_\Sigma^{} \bigr) }   
- \frac{ (D-F)\, \bigl( h_D^{}-3 h_F^{} \bigr) 
  }{  \sqrt{6}\,\, \bigl( m_\Xi^{}-m_\Lambda^{} \bigr) }   \,\,, 
\vspace{2ex} \\   \displaystyle    
{a}^{(P)}_{\Sigma^+ n\pi^+}  \,=\,   \displaystyle  
\frac{-D\, \bigl( h_D^{}-h_F^{} \bigr) }{ m_\Sigma^{}-m_N^{}}   
- \frac{ \mbox{$\frac{1}{3}$} D\, \bigl( h_D^{}+3 h_F^{} \bigr) 
  }{  m_\Lambda^{}-m_N^{} }   \,\,, 
\vspace{2ex} \\   \displaystyle    
{a}^{(P)}_{\Sigma^- n\pi^-}  \,=\,   \displaystyle  
\frac{-F\, \bigl( h_D^{}-h_F^{} \bigr) }{ m_\Sigma^{}-m_N^{}}   
- \frac{ \mbox{$\frac{1}{3}$} D\, \bigl( h_D^{}+3 h_F^{} \bigr) 
  }{  m_\Lambda^{}-m_N^{} }   \,\,.     
\end{array} 
\end{eqnarray}    
\label{cptamp}   
\end{subequations}   \\
The leading nonanalytic contributions to the amplitudes arise 
from one-loop diagrams, with $h_C^{}$ only appearing in those 
involving decuplet baryons.  
These contributions have been calculated by various 
authors~\cite{bsw,jenkins2,nlhd,at}, and we will adopt the results 
of Ref.~\cite{at} for the numerical estimate of our uncertainty.

\begin{figure}[ht]         
\includegraphics{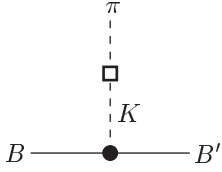}   \vspace{-3ex}
\caption{\label{tree1}%
Kaon-pole diagram contributing to $P$-wave hyperon 
nonleptonic decays.
The weak vertex here comes from the  $\gamma_8^{}$  term in  
Eq.~(\ref{weakcl})} 
\end{figure}

In Fig.~\ref{tree1}, we show the kaon-pole diagram to be 
discussed later on. 
In this diagram, there is a strong vertex from Eq.~(\ref{Ls1}) 
followed by a kaon pole and a weak vertex from the $\gamma_8^{}$ 
term in Eq.~(\ref{weakcl}).   
Notice that this term is not only subleading in the chiral 
expansion, but also suppressed by an $m_\pi^2/m_K^2$  factor 
(and hence vanishing in the  $\,m_u^{}=m_d^{}=0\,$  limit).

Once the values of the weak couplings $h_{D,F}^{}$ are specified,  
the formulas in Eq.~(\ref{cptamp}) determine the leading-order 
amplitudes.  
It is well known that this representation does not provide a good 
fit to the measured $P$-wave amplitudes, and that higher-order 
terms are important~\cite{bsw,dgh,jenkins2,nlhd,at,BorHol}. 
The procedure that we adopt for estimating the weak phases is to 
obtain the real part of the amplitudes from experiment (assuming 
no $CP$ violation) and to use Eq.~(\ref{cptamp}) to estimate 
the imaginary parts.  
The dominant $CP$-violating phases in the $\,|\Delta I|=\frac{1}{2}\,$  
sector of the $\,|\Delta S|=1\,$  weak interaction occur in the Wilson 
coefficient  $C_6^{}$  associated with the penguin operator $Q_6^{}$.  
Our strategy will be to calculate within a model the imaginary part 
of the couplings  $h_{D,F,C}^{}$ and $\gamma_8^{}$  induced by $Q_6^{}$. 
As a numerical result, we propose a central value from 
leading-order $\chi$PT [Eq.~(\ref{cptamp})] and an estimate of the 
error from the nonanalytic corrections obtained with the 
expressions given in Ref.~\cite{at}.

To end this section, for later use we collect in Table~\ref{spx} 
the experimental values of the $S$- and $P$-wave amplitudes of 
interest, reproduced from Ref.~\cite{at}.  
The numbers are extracted (neglecting strong and weak phases) 
from the measured decay width  $\Gamma$  and decay parameter  $\alpha$ 
by means of the relations  
\begin{eqnarray}   \label{a,b,c} 
\begin{array}{c}   \displaystyle   
\Gamma  \,=\,   
\frac{ \bigl| \mbox{$\bm{p}$}_{B'}^{} \bigr| }{4\pi\, m_{B}^{}} 
\bigl( E_{B'}^{}+m_{B'}^{} \bigr) 
\left( |s|^2 + |p|^2 \right)   \,\,,   
\hspace{3em}  
\alpha  \,=\,  \frac{2\,{\rm Re}(s^*p)}{|s|^2 + |p|^2}   \,\,.   
\end{array}       
\end{eqnarray}       
The $s$ and  $p$ amplitudes are related to those in 
Eq.~(\ref{M,HB}) above by\footnote{In Refs.~\cite{jenkins2,at},  
the $p$ expression has the opposite sign, 
$\,p=-|\mbox{$\bm{p}$}_{B'}^{}|\, {\cal A}^{(P)},\,$  
but this turns out to be inconsistent with the amplitude formula 
from which both $\Gamma$ and $\alpha$ are derived.   
Nevertheless, the sign flip does not affect the conclusions 
of Refs.~\cite{jenkins2,at}, as the fits therein were performed 
to the $S$-waves and the $P$-waves were poorly reproduced 
regardless of the sign of $p$.
}    
\begin{eqnarray}   \label{sp,HB} 
s  \,=\,  {\cal A}^{(S)}     \,\,,   
\hspace{3em}  
p  \,=\,  \bigl| \mbox{$\bm{p}$}_{B'}^{} \bigr| \, {\cal A}^{(P)}   \,\,.
\end{eqnarray}    
\begin{table}[ht]      
\caption{\label{spx}%
Experimental values for $S$- and $P$-wave amplitudes, in units 
of  $G_{\rm F}^{}m_{\pi^+}^2$.}      
\centering   \footnotesize 
\vskip 0.5\baselineskip
\begin{tabular}{c||c|c}    
\hline \hline      
\raisebox{-0.5ex}{Decay mode $\vphantom{\Bigl|}$}   &    
\raisebox{-0.5ex}{$\hspace{1ex}s$}   &
\raisebox{-0.5ex}{$\hspace{1ex}p$}   
\\ \hline && \vspace{-3ex} \\           
$\vphantom{\begin{array}{c}\vspace{9ex}\end{array}}$  
$\begin{array}{rcl}   \displaystyle  
\Lambda   & \hspace{-0.5em} \to & \hspace{-0.5em}  p\pi^-    
\\        
\Xi^-     & \hspace{-0.5em} \to & \hspace{-0.5em}  \Lambda\pi^- 
\\  
\Sigma^+  & \hspace{-0.5em} \to & \hspace{-0.5em}  n\pi^+ 
\\      
\Sigma^+  & \hspace{-0.5em} \to & \hspace{-0.5em}  p\pi^0 
\\      
\Sigma^-  & \hspace{-0.5em} \to & \hspace{-0.5em}  n\pi^-   
\end{array}$  
&  
$\begin{array}{rcl}   \displaystyle  
 1.42  & \hspace{-.5em} \pm & \hspace{-.5em}  0.01   \\  
-1.98  & \hspace{-.5em} \pm & \hspace{-.5em}  0.01   \\ 
 0.06  & \hspace{-.5em} \pm & \hspace{-.5em}  0.01   \\  
-1.43  & \hspace{-.5em} \pm & \hspace{-.5em}  0.05   \\  
 1.88  & \hspace{-.5em} \pm & \hspace{-.5em}  0.01   
\end{array}$  
&  
$\begin{array}{rcl}   \displaystyle   
 0.52  & \hspace{-.5em} \pm & \hspace{-.5em}  0.01   \\  
 0.48  & \hspace{-.5em} \pm & \hspace{-.5em}  0.02   \\ 
 1.81  & \hspace{-.5em} \pm & \hspace{-.5em}  0.01   \\  
 1.17  & \hspace{-.5em} \pm & \hspace{-.5em}  0.06   \\  
-0.06  & \hspace{-.5em} \pm & \hspace{-.5em}  0.01   
\end{array}$ 
\\   \hline \hline 
\end{tabular}   
\end{table}

\section{Estimate of counterterms\label{ect}}
        
Our task in this section is to match the dominant  
$\,|\Delta I|=\frac{1}{2}\,$  $CP$-violating term from the 
standard-model effective weak Hamiltonian in Eq.~(\ref{weaksd}) 
to the weak chiral Lagrangian in Eq.~(\ref{weakcl}).  
That is, to compute the imaginary part of the parameters 
$h_D^{}$,  $h_F^{}$,  $h_C^{}$, and $\gamma_8^{}$ that is induced 
by  $\,{\rm Im}\, C_6^{} Q_6^{}\,$  in Eq.~(\ref{weaksd}).  
To do this, we will include both factorizable contributions, 
that arise from regarding the operator $Q_6^{}$ as the product of two 
(pseudo)scalar densities, and direct (nonfactorizable) 
contributions calculated in the MIT bag model.

The nonfactorizable contributions are easily obtained from the 
observation that the weak chiral Lagrangian of Eq.~(\ref{weakcl}) 
is responsible for nondiagonal ``weak mass terms'' such as  
\begin{eqnarray}   \label{<B'|H|B>}   
\begin{array}{c}   \displaystyle   
\bigl\langle n \bigr| \bigl( {\cal H}_{\rm w}^{} \bigr) _8^{}
\bigl| \Lambda \bigr\rangle  
\,=\,  
\frac{h_D^{}+3 h_F^{}}{\sqrt 6}\, \bar u_n^{} u_\Lambda^{}  \,\,,  
\vspace{2ex} \\   \displaystyle    
\bigl\langle \Lambda \bigr| \bigl( {\cal H}_{\rm w}^{} \bigr) _8^{}
\bigl| \Xi^0 \bigr\rangle  
\,=\,  
\frac{h_D^{} - 3 h_F^{}}{\sqrt 6}\, \bar u_\Lambda^{} u_\Xi^{}   \,\,,  
\vspace{2ex} \\   \displaystyle    
\bigl\langle \Xi^{*-} \bigr| \bigl( {\cal H}_{\rm w}^{} \bigr) _8^{}  
\bigl| \Omega^- \bigr\rangle  
\,=\,  \frac{-h_C^{}}{\sqrt 3}\, \bar u_{\Xi^*}^{}\cdot u_\Omega^{}  \,\,,    
\end{array}      
\end{eqnarray}      
where the subscript 8 denotes the component of  ${\cal H}_{\rm w}^{}$  
that transforms as  $(8_{\rm L}^{},1_{\rm R}^{})$.  
These terms can be computed directly from the short-distance 
Hamiltonian in Eq.~(\ref{weaksd}) by calculating  in the MIT bag 
model the baryon-baryon matrix elements of the four-quark operators. 
From the basic results in Appendix~\ref{bagmodel}, 
one finds the  $Q_6^{}$ contributions  
\begin{eqnarray}   \label{hd,hf,hc}  
\begin{array}{c}   \displaystyle   
h_D^{}  \,=\,   
\frac{G_{\rm F}^{} \lambda}{\sqrt 2}\, C_6^{}\, (3\, a-5\, b)   \,\,,  
\hspace{2em}  
h_F^{}  \,=\,  
\frac{G_{\rm F}^{} \lambda}{\sqrt 2}\, C_6^{}\, 
\Bigl( a+\mbox{$\frac{11}{3}$}\, b \Bigr)   \,\,,   
\vspace{2ex} \\   \displaystyle    
h_C^{}  \,=\,  
\frac{G_{\rm F}^{} \lambda}{\sqrt 2}\, C_6^{}\, (-12\, a+4\, b)   \,\,,    
\end{array}      
\end{eqnarray}      
where  $a$  and  $b$  are bag parameters whose values are given in 
Eq.~(\ref{ob}) for  $h_{D,F}^{}$  and in  Eq.~(\ref{db})  for  $h_C^{}$.   
Numerically, the imaginary part of $C_6^{}$ then yields, in units of  
$\,\sqrt 2\, f_\pi^{} G_{\rm F}^{} m_{\pi^+}^2\, \lambda^5 A^2\eta,\,$   
\begin{eqnarray}   \label{Imh[bag]}  
\begin{array}{c}   \displaystyle   
{\rm Im}\, h_D^{}  \,=\,  0.278 \,\, y_6^{}   \,\,,  
\hspace{2em}  
{\rm Im}\, h_F^{}  \,=\,  1.04 \,\,  y_6^{}   \,\,,  
\hspace{2em}  
{\rm Im}\, h_C^{}  \,=\,  -3.13 \,\, y_6^{}   \,\,,  
\end{array}      
\end{eqnarray}      
where  $\,f_\pi^{}\simeq 92.4\,\rm MeV\,$  has been used.  
The units are chosen to separate both the conventional normalization 
for the hyperon decay amplitudes, as in Eq.~(\ref{conamp}) and 
Table~\ref{spx}, and the relevant combination of CKM parameters 
occurring in the observables $A_{\Lambda,\Xi}^{}$.

To obtain the factorizable contributions to the imaginary part of 
the parameters  $h_{D,F,C}^{}$,  we follow the procedure used in 
kaon physics for  $\gamma_8^{}\,$~\cite{penfac}.  
As shown in Appendix~\ref{wpfac}, the lowest-order chiral 
realization of a factorized $Q_6^{}$ contributes to the weak 
Lagrangian in Eq.~(\ref{weakcl}) with
\begin{eqnarray}   \label{h[fac]}
\begin{array}{c}   \displaystyle   
h_D^{}  \,=\,  
\frac{G_{\rm F}^{}\, \lambda}{ \sqrt 2}\,\, 
8\, C_6^{}\, f^2 B_0^{}\, b_D^{}   \,\,,  
\hspace{3em}  
h_F^{}  \,=\,  
\frac{G_{\rm F}^{}\, \lambda}{\sqrt 2}\,\, 
8\, C_6^{}\, f^2 B_0^{}\, b_F^{}   \,\,,  
\vspace{2ex} \\   \displaystyle    
h_C^{}  \,=\,  
\frac{G_{\rm F}^{}\, \lambda}{\sqrt 2}\,\, 
8\, C_6^{}\, f^2 B_0^{}\, c   \,\,,  
\hspace{3em}  
\gamma_8^{}  \,=\,  
\frac{G_{\rm F}^{}\, \lambda}{\sqrt 2}\,\, 
16\, C_6^{}\, B_0^2\, L_5^{}   \,\,.  
\end{array}      
\end{eqnarray}      
The values of  $b_D^{}$, $b_F^{}$, and $c$ can be determined by 
fitting the mass formulas derived from the Lagrangian in 
Eq.~(\ref{Ls2}), with $\chi=2B_0^{} M,\,$  to the measured 
masses of the octet and decuplet baryons~\cite{pdb}.       
Thus we find   
\begin{eqnarray}   \label{bDbF}
b_D^{}\, m_s^{}  \,=\,  0.0301\,\,{\rm GeV}   \,\,,  
\hspace{2em}  
b_F^{}\, m_s^{}  \,=\,  -0.0948\,\,{\rm GeV}   \,\,,  
\hspace{2em}  
c\, m_s^{}  \,=\,  0.221\,\,{\rm GeV}   \,\,
\end{eqnarray}      
for  $\,m_u^{}=m_d^{}=0.\,$    
In this limit, the Lagrangian in Eq.~(\ref{Ls2}) also gives  
$\,m_K^2=B_0^{}\, m_s.\,$   
Using  $\,m_s^{}=\bar m_s^{}(\mu=m_c^{})=170\,\rm MeV\,$  
from  Ref.~\cite{buras}, we then have    
\begin{eqnarray}   \label{bDbFtree}
b_D^{}  \,=\,  0.177   \,\,,  \hspace{2em}  
b_F^{}  \,=\,  -0.558  \,\,,   \hspace{2em}   
c  \,=\,  1.30  \,\,,   \hspace{2em}   
B_0^{}  \,=\,  1.45~{\rm GeV}  \,\,.   
\end{eqnarray}      
For  $L_5^{}$,  we adopt the value  $\,L_5^{}=1.4\times10^{-3}\,$  
found in Ref.~\cite{gl}. 
Setting  $\,f=f_\pi^{}\simeq92.4\,\rm MeV,\,$  
we then obtain the  $Q_6^{}$  contributions,  in units of
$\,\sqrt 2\, f_\pi^{} G_{\rm F}^{} m_{\pi^+}^2\, 
\eta \lambda^5 A^2,\,$      
\begin{eqnarray}   \label{Imh[fac]}
\begin{array}{c}   \displaystyle   
{\rm Im}\, h_D^{}  \,=\,  4.86\, y_6^{}   \,\,,  
\hspace{3em}  
{\rm Im}\, h_F^{}  \,=\,  -15.3\, y_6^{}   \,\,,  
\hspace{3em}  
{\rm Im}\, h_C^{}  \,=\,  35.6\, y_6^{}   \,\,,   
\vspace{2ex} \\   \displaystyle    
{\rm Im}\, \gamma_8^{}\; B_0^{}  \,=\,  
18.8\, y_6^{} B_6^{(1/2)}   \,\,,  
\end{array}      
\end{eqnarray}      
where the formula with  $\gamma_8^{}$  is the usual one appearing 
in the calculation of  $\epsilon^\prime$ in kaon decay, and we have 
introduced the standard parameter  $B_6^{(1/2)}\,$  to encode 
deviations from factorization~\cite{buras}, so that here  
$\,B_6^{(1/2)}=1.\,$

\section{Numerical Results\label{results}}

If Eq.~(\ref{cptamp}) provided a good fit to the hyperon decay 
amplitudes, it would be straightforward to calculate the weak phases 
of  Eq.~(\ref{approxa}).   
We would simply divide the imaginary part of the amplitudes by 
the real part of the amplitudes obtained from a matching of 
the parameters $h_{D,F}^{}$ to the short-distance Hamiltonian.  
However, as we mentioned before, leading-order chiral perturbation 
theory fails to reproduce simultaneously the $S$- and $P$-wave 
amplitudes. 
Consequently, we are forced to employ the real part of the amplitudes 
that are extracted from experiment under the assumption of no $CP$ 
violation.

An additional problem occurs if we calculate the imaginary part of the 
amplitudes from a matching of the {\it full} weak Hamiltonian to 
$h_{D,F}^{}$  and then divide it by the {\it experimental} amplitudes, 
as this introduces spurious phase differences. This can 
be easily understood by considering the case where only one operator 
occurs in the short-distance weak Hamiltonian. 
In such a case, it is clear that there can be no $CP$ violation, as 
there is only one weak phase in the problem. However, if we use 
the procedure outlined above to calculate the phase difference 
$\phi_S^{}-\phi_P^{}$, we obtain a nonzero result due to the mismatch 
between the predicted and the measured ratio $p/s$.

On the other hand, if there are two operators in the short-distance 
weak Hamiltonian, and one of them is mostly responsible for the 
real part of the amplitudes while the other one is mostly responsible 
for the weak phases, the procedure above does not introduce spurious 
phases. Of course, the predictions obtained are reliable only to the 
extent that the model reproduces the true imaginary part of the amplitudes.

In view of all this, we adopt the following prescription to obtain 
the weak phases.   
We first assume that the real part of the weak decay amplitudes  
originates predominantly in the tree-level operators $Q_{1,2}$.  
This is true in the bag model, for example,  as can be seen from 
the results in Appendix~\ref{bagmodel}. 
We then assume that the imaginary part of the amplitudes is primarily 
due to the  $\,{\rm Im}\, C_6^{} Q_6^{}\,$ term in the weak Hamiltonian.  
This is true both in the bag model and in the vacuum-saturation 
model of Ref.~\cite{hsv}, and is due to the purely  
$\,|\Delta I|=\frac{1}{2}\,$  nature of the $CP$ observables 
$A_{\Lambda,\Xi}^{}$.  
With these assumptions,  we calculate a central value for the 
imaginary part of the weak decay amplitudes using Eq.~(\ref{cptamp}) 
with values for  ${\rm Im}\, h_{D,F}^{}$  obtained in the previous 
section by adding the factorizable and nonfactorizable contributions.   
We estimate the uncertainty in this prediction by computing   
the leading nonanalytic corrections with our values for 
${\rm Im}\,h_{D,F,C}^{}$.\footnote{This prescription of taking 
the leading nonanalytic contributions as the uncertainty in 
the lowest-order amplitudes works remarkably well for the real 
part of the amplitudes.  
To show this, we use the weak parameters determined from fitting 
simultaneously  the  $S$-wave amplitudes in Eq.~(\ref{stree}) and 
the leading-order $P$-wave amplitudes for  
$\,\Omega\to\Lambda K,\Xi\pi\,$  provided by Ref.~\cite{jenkins2}  
to the measured amplitudes.  
Thus, $\,h_D^{}=0.49,\,$  $\,h_F^{}=1.18,\,$  and  
$\,h_C^{}=1.15,\,$  all in units of 
$\,\sqrt 2\, f_\pi^{} G_{\rm F}^{} m_{\pi^+}^2.\,$    
Writing the resulting amplitudes as tree$\pm$loop, and excluding 
$\gamma_8^{}$  terms,  we have 
$\,s_{\Lambda\to p\pi^-}^{}=1.25\pm2.28,\,$  
$\,s_{\Xi^-\to\Lambda\pi^-}^{}=-1.65\pm2.96,\,$  
$\,p_{\Lambda\to p\pi^-}^{}=0.49\pm0.92,\,$  
and  $\,p_{\Xi^-\to\Lambda\pi^-}^{}=-0.16\pm2.21,\,$  
all in units of $G_{\rm F}^{} m_{\pi^+}^2$. 
Clearly the corresponding data in Table~\ref{spx} are 
well within these ranges.   
}

In order to compare with older results in the literature, we have 
calculated two additional terms, both proportional to $\gamma_8^{}$, 
in which the $CP$-violating weak transition occurs in the meson sector. 
The tree-level kaon-pole contribution to the $P$-waves will be shown 
in one of our tables because this is in fact the dominant 
contribution to the commonly quoted result of Donoghue, He, and 
Pakvasa~\cite{dhp}, as we discuss below. 
The one-loop nonanalytic contribution proportional to  $\gamma_8^{}$  
occurs  at order $p^3$ in the chiral expansion.   
It is related to the model employed by Iqbal and Miller in  
Ref.~\cite{im}, and we include it here to comment on that 
result.

For our numerical calculations, we use the leading-order 
(in QCD) Wilson coefficients at $\,\mu=m_c^{}=1.3\,\rm GeV\,$   
listed in Table XIX of Ref.~\cite{buras}.   
In particular,   
\begin{eqnarray}   \label{y6}
y_6^{}  \,=\,  -0.096   \,\,,   
\end{eqnarray}      
corresponding to  
$\,\Lambda_{\overline{\rm MS}}^{(4)}=325\,\rm MeV.\,$  
This is one of the middle values of  $y_6^{}$  in this table,  
which vary from  $-0.063$  to  $-0.120$,  depending on 
the value of  $\Lambda_{\overline{\rm MS}}^{(4)}$  and 
on the renormalization scheme.  
In the rest of this section, we numerically evaluate the weak 
phases in the $\Lambda$ and $\Xi^-$  decays, relegating 
the corresponding evaluation for the $\Sigma$ decays to 
Appendix~\ref{SigmaNpi}.

The nonfactorizable contributions from $Q_6^{}$ to the weak 
parameters are given by the bag-model results in  Eq.~(\ref{Imh[bag]}).   
The resulting  $s$ and $p$  amplitudes  are collected 
in Table~\ref{spbag}, divided by 
the experimental amplitudes of Table~\ref{spx}. 
For the factorizable contributions, the parameters are given in 
Eq.~(\ref{Imh[fac]}) and the corresponding amplitudes are 
listed in Table~\ref{spfac}.  
In calculating the imaginary parts in these tables, 
we employ the  $y_6^{}$  value in Eq.~(\ref{y6}), as well as 
the strong couplings   $\,D=0.8,\,$  $\,F=0.5,\,$  
$\,|{\cal C}|=1.7,\,$  and  $\,{\cal H}=-2.4.\,$  
The loop contributions are computed using the results 
of Ref.~\cite{at} at a renormalization scale of  $1\,\rm GeV$  
with  $\,f_P^{}=f_\pi^{},\,$  and serve as an error estimate of the 
prediction given by the tree contributions.   
In Table~\ref{spfac}, we have separated out the terms 
containing ${\rm Im}\, \gamma_8^{}$.  
In the $P$-waves, the $\gamma_8^{}$ contributions also occur 
at next-to-leading tree-level order, arising from the kaon-pole 
diagram in Fig.~\ref{tree1}.

\begin{table}[t]   
\caption{\label{spbag}%
Ratios of the imaginary part of the theoretical value to the 
experimental value, for $S$- and $P$-wave amplitudes, with the weak 
couplings from  $Q_6^{}$ contribution only, estimated in the bag model.     
The ratios are in units of  $\eta \lambda^5 A^2$.
}   
\centering   \footnotesize
\vskip 0.5\baselineskip
\begin{tabular}{c||cc|cc}   
\hline \hline 
Decay mode $\vphantom{\Biggl|_o^o}$  &  
$\displaystyle \frac{{\rm Im}\,s_{\rm tree}^{}}{s_{\rm expt}^{}}$ &
$\hspace{1ex}\displaystyle
\frac{{\rm Im}\,s_{\rm loop}^{}}{s_{\rm expt}^{}}$ &
$\hspace{1ex}\displaystyle 
\frac{{\rm Im}\,p_{\rm tree}^{}}{p_{\rm expt}^{}}$ & 
$\hspace{1ex}\displaystyle  
\frac{{\rm Im}\,p_{\rm loop}^{}}{p_{\rm expt}^{}}$  
\\ \hline && && \vspace{-3ex} \\      
$\vphantom{\begin{array}{c}\vspace{4ex}\end{array}}$  
$\begin{array}{rcl}   \displaystyle
\Lambda   & \hspace{-.5em} \to & \hspace{-.5em} p\pi^- \\     
\Xi^-     & \hspace{-.5em} \to & \hspace{-.5em} \Lambda\pi^- 
\end{array}$ 
&  
$\begin{array}{r}   \displaystyle
-0.09 \\ -0.06  \end{array}$  &
$\begin{array}{r}   \displaystyle
-0.09 \\ -0.06  \end{array}$  &   
$\begin{array}{r}   \displaystyle 
-0.15 \\  0.14  \end{array}$  &
$\begin{array}{r}   \displaystyle
 0.28 \\ -0.26  \end{array}$ 
\\
\hline \hline  
\end{tabular}    
%
\bigskip  
%
\caption{\label{spfac}%
Ratios of the imaginary part of the theoretical value to the 
experimental value, for $S$- and $P$-wave amplitudes, with the weak 
couplings from  $Q_6^{}$ contribution only, estimated in factorization.     
The ratios are in units of  $\eta \lambda^5 A^2$.    
}  
\centering   \footnotesize
\vskip 0.5\baselineskip
\begin{tabular}{c||ccc|cccc}   
\hline \hline 
Decay mode $\vphantom{\Bigg|_{(o)}^{(o)}}$       &  
$\displaystyle \frac{{\rm Im}\,s_{\rm tree}^{}}{ s_{\rm expt}^{}}$ &
$\hspace{1ex}\displaystyle
\frac{{\rm Im}\,s_{\rm loop}^{}}{ s_{\rm expt}^{}}$ &
$\hspace{1ex}\displaystyle
\frac{{\rm Im}\,s_{\rm loop}^{(\gamma_8^{})}}{ s_{\rm expt}^{}}$ &
$\hspace{1ex}\displaystyle 
\frac{{\rm Im}\,p_{\rm tree}^{}}{ p_{\rm expt}^{}}$ & 
$\hspace{1ex}\displaystyle  
\frac{{\rm Im}\,p_{\rm tree}^{(\gamma_8^{})}}{ p_{\rm expt}^{}}$ & 
$\hspace{1ex}\displaystyle  
\frac{{\rm Im}\,p_{\rm loop}^{}}{ p_{\rm expt}^{}}$ & 
$\hspace{1ex}\displaystyle  
\frac{{\rm Im}\,p_{\rm loop}^{(\gamma_8^{})}}{ p_{\rm expt}^{}}$   
\\ \hline && && && & \vspace{-3ex} \\      
$\vphantom{\begin{array}{c}\vspace{4ex}\end{array}}$  
$\begin{array}{rcl}   \displaystyle  
\Lambda   & \hspace{-.5em} \to & \hspace{-.5em} p\pi^- \\     
\Xi^-     & \hspace{-.5em} \to & \hspace{-.5em} \Lambda\pi^- 
\end{array}$
&
$\begin{array}{r}   \displaystyle
 1.13 \\  1.00  \end{array}$  &
$\begin{array}{r}   \displaystyle
 1.05 \\  1.10  \end{array}$  &   
$\begin{array}{r}   \displaystyle
 0.35 \\  0.56  \end{array}$   
&  
$\begin{array}{r}   \displaystyle 
 1.33 \\ -0.66  \end{array}$  &
$\begin{array}{r}   \displaystyle 
 0.04 \\ -0.02  \end{array}$  &
$\begin{array}{r}   \displaystyle
-0.27 \\  0.59  \end{array}$  &
$\begin{array}{r}   \displaystyle
 0.61 \\ -0.28  \end{array}$
\\   
\hline \hline  
\end{tabular}
\bigskip  
\end{table}

In Table~\ref{phases}, we combine the weak phases from the preceding 
two tables, keeping  only the leading-order and loop contributions 
(excluding $\gamma_8^{}$ terms). 
We also show in this table another error estimate, $\delta\phi$,  
obtained from the leading-order amplitudes, but allowing 
the parameters to vary between their tree-level and 
one-loop values.      
In making this estimate, we use only the factorization amplitudes, 
as they are are much larger than the bag-model contributions, 
as seen in the previous two tables.   
Thus, for the $S$-wave amplitudes, we need the one-loop values of
the parameters  $b_{D,F}^{}$.     
Employing the one-loop formulas for baryon masses derived in 
Ref.~\cite{jenkins1}, we find 
\begin{eqnarray}   \label{bDbFloop}
b_D^{}  \,=\,  -0.636   \,\,,   \hspace{3em}     
b_F^{}  \,=\,  -0.192   \,\,.  
\end{eqnarray}     
For the $P$-waves, we note that the factorization 
parameters in Eq.~(\ref{h[fac]})  and  the tree-level 
mass formulas
\begin{eqnarray}  
\begin{array}{c}   \displaystyle   
m_\Sigma^{} - m_N^{}  \,=\,  2\bigl(b_D^{}-b_F^{}\bigr)\, m_s^{}   \,\,,  
\hspace{3em}  
m_\Lambda^{} - m_N^{}  \,=\,  
-\mbox{$\frac{2}{3}$} \bigl( b_D^{}+3 b_F^{}\bigr) \, m_s^{}   \,\,,   
\vspace{2ex} \\   \displaystyle    
m_\Xi^{} - m_\Sigma^{}  \,=\,  
-2 \bigl(b_D^{}+b_F^{}\bigr)\, m_s^{}   \,\,,
\hspace{3em}  
m_\Xi^{} - m_\Lambda^{}  \,=\,  
\mbox{$\frac{2}{3}$} \bigl( b_D^{}-3 b_F^{} \bigr) \, m_s^{}   \,\,,     
\end{array}      
\end{eqnarray}      
derived from Eq.~(\ref{Ls2}), lead to simplified expressions for 
the leading-order amplitudes arising from the $Q_6^{}$ contribution, 
namely,  
\begin{eqnarray}    \label{Pfac}    
\begin{array}{c}   \displaystyle  
{a}^{(P)}_{\Lambda p\pi^-}  \,=\,   \displaystyle      
\frac{G_{\rm F}^{}\, \lambda}{ \sqrt 2}\, 
\frac{4\, C_6^{}\, f^2 B_0^{}}{\sqrt 6\, m_s^{}}\, (-D-3F)   \,\,,   
\hspace{3em} 
{a}^{(P)}_{\Xi^-\Lambda\pi^-}  \,=\,   \displaystyle   
\frac{G_{\rm F}^{}\, \lambda}{ \sqrt 2}\, 
\frac{4\, C_6^{}\, f^2 B_0^{}}{\sqrt 6\, m_s^{}}\, (-D+3F)   \,\,,     
\vspace{2ex} \\   \displaystyle    
{a}^{(P)}_{\Sigma^+ n\pi^+}  \,=\,   0   \,\,,  
\hspace{3em}  
{a}^{(P)}_{\Sigma^- n\pi^-}  \,=\,   \displaystyle      
\frac{G_{\rm F}^{}\, \lambda}{ \sqrt 2}\, 
\frac{4\, C_6^{}\, f^2 B_0^{}}{ m_s^{}}\, (D-F)   \,\,,   
\end{array} 
\end{eqnarray}    
where the  $\Sigma$-decay amplitudes have been included to be used  
in  Appendix~\ref{SigmaNpi}.  
Consequently, we only need the one-loop values of $D$  and  $F$.  
A one-loop fit to the semileptonic hyperon decays yields~\cite{JenMan3}    
\begin{eqnarray}   
D  \,=\,  0.61  \,\,,   \hspace{3em}     
F  \,=\,  0.40  \,\,.  
\label{sploop}   
\end{eqnarray}     
Using these results, together with their tree-level counterparts 
in Eqs.~(\ref{sptree})  and~(\ref{bDbFtree}), we write the ranges   
\begin{eqnarray}   
\begin{array}{c}   \displaystyle   
-0.64  \,\le\,  b_D^{}  \,\le\,  +0.18   \,\,,   \hspace{3em}     
-0.56  \,\le\,  b_F^{}  \,\le\,  -0.19   \,\,,   
\vspace{2ex} \\   \displaystyle    
0.61  \,\le\,  D  \,\le\,  0.80  \,\,,   \hspace{3em}     
0.40  \,\le\,  F  \,\le\,  0.50  \,\,.  
\end{array}     
\end{eqnarray}     
We take  $\delta\phi_{S,P}^{}$  to be the largest deviation  
from  $\phi_{S,P}^{\rm tree}$  (in factorization) allowed 
by these ranges.

\begin{table}[t]   
\caption{\label{phases}%
Weak $S$- and  $P$-wave phases from  $Q_6^{}$ contribution alone,   
in units of  $\eta \lambda^5 A^2$.    
}  
\centering   \footnotesize
\vskip 0.5\baselineskip    
\begin{tabular}{c||ccc|ccc}
\hline \hline 
Decay mode $\vphantom{\bigg|}$  &  
$\,\,\phi_S^{\rm(tree)}\,\,$  &  $\,\,\phi_S^{\rm(loop)}\,\,$  &
$\,\,\delta\phi_S^{\rm(tree)}\,\,$  &   
$\,\,\phi_P^{\rm(tree)}\,\,$  &  $\,\,\phi_P^{\rm(loop)}\,\,$  &
$\,\,\delta\phi_P^{\rm(tree)}\,\,$  
\\ \hline && && && \vspace{-3ex} \\     
$\vphantom{\begin{array}{c} \vspace{1ex} \end{array}}$ 
$\begin{array}{rcl}   \displaystyle
\Lambda   & \hspace{-.5em} \to & \hspace{-.5em} p\pi^- \\     
\Xi^-     & \hspace{-.5em} \to & \hspace{-.5em} \Lambda\pi^-
\end{array}$
&   
$\begin{array}{r}  1.04 \\  0.94  \end{array}$  &
$\begin{array}{r}  0.96 \\  1.04  \end{array}$  &   
$\begin{array}{r} -0.83 \\ -1.04  \end{array}$  &  
$\begin{array}{r}  1.18 \\ -0.52  \end{array}$  &
$\begin{array}{r}  0.01 \\  0.33  \end{array}$  &
$\begin{array}{r} -0.30 \\  0.27  \end{array}$  
\\   \hline \hline  
\end{tabular}   
\bigskip  
\end{table}

From the numbers in Table~\ref{phases}, we may conclude that the 
uncertainties of  $\phi_S^{}$  and  $\phi_P^{}$  are of order  100$\%$  
and  50$\%$, respectively,  for both decays.    
This is reflected in our prediction for the phases, which are collected  
in Table~\ref{dphases} along with the resulting phase differences.     
The errors for the differences have been obtained simply by adding 
the individual errors.  
We have also collected strong-phase differences in the table, 
from the numbers given in the Introduction.  
The errors we quote in this table are obviously not Gaussian. 
They simply indicate the allowed ranges within our prescription to 
calculate the phases.

\begin{table}[ht]   
\caption{\label{dphases}%
Weak phases in units of  $\eta \lambda^5 A^2$,  
and strong-phase differences,  $\delta_S^{}-\delta_P^{}$.    
}  
\centering   \footnotesize
\vskip 0.5\baselineskip    
\begin{tabular}{c||ccc|c}  
\hline \hline
Decay mode $\vphantom{\bigg|}$  &  $\phi_S^{}$  &  
\hspace{1em}$\phi_P^{}$  &  $\phi_S^{}-\phi_P^{}$  & 
$\delta_S^{}-\delta_P^{}$  
\\ \hline && && \vspace{-3ex} \\   
$\vphantom{\begin{array}{c}\\ \vspace{1ex} \end{array}}$ 
$\begin{array}{rcl}   \displaystyle
\Lambda   & \hspace{-.5em} \to & \hspace{-.5em} p\pi^- \\     
\Xi^-     & \hspace{-.5em} \to & \hspace{-.5em} \Lambda\pi^-
\end{array}$  &  
$\begin{array}{c}   \displaystyle
1.0 \pm 1.0 \\  0.9\pm0.9  \end{array}$  &    
$\begin{array}{r}   \displaystyle
1.2 \pm 0.6 \\  -0.5\pm0.3 \end{array}$  &    
$\begin{array}{c}   \displaystyle
-0.2 \pm 1.6 \hspace{1.6ex} \\  1.4\pm 1.2  \end{array}$  &    
$\begin{array}{c}   \displaystyle
7^\circ \pm 2^\circ \\  1.1^\circ\pm2.8^\circ  \end{array}$  
\\ \hline \hline  
\end{tabular}
\bigskip  
\end{table}

Putting together these results, we finally obtain
\begin{eqnarray}   \label{A}
\begin{array}{c}   \displaystyle  
A(\Lambda^0_-)  \,=\,  A_\Lambda^{}  \,=\,  
(0.03\pm0.25)\,\, A^2\lambda^5\eta   \,\,,   
\vspace{2ex} \\   \displaystyle    
A(\Xi^-_-)  \,=\,  A_\Xi^{}  \,=\,  
(-0.05\pm0.13)\,\, A^2\lambda^5\eta    \,\,,   
\end{array}
\end{eqnarray}
leading to  
\begin{equation}
A_{\Xi\Lambda}^{}  \,=\,  
A_\Lambda^{} + A_\Xi^{}   
\,=\,  (-0.02\pm0.38)\,\, A^2\lambda^5\eta   \,\,.       
\end{equation}
With the CKM parameter values given in Eq.~(\ref{ckm}),   
we have   $\,A^2\lambda^5\eta\simeq1.26\times10^{-4}\,$   
and  therefore
\begin{eqnarray}
\begin{array}{c}   \displaystyle  
-3\times10^{-5}  \,\le\,  A_\Lambda^{}  \,\le\,  4\times10^{-5}  \,\,,   
\hspace{3em}  
-2\times10^{-5}  \,\le\,  A_\Xi^{}  \,\le\,  1\times10^{-5}  \,\,,   
\end{array}
\end{eqnarray}
\begin{equation}   
-5\times 10^{-5}  \,\le\,  A_{\Xi\Lambda}^{}  \,\le\,
5\times 10^{-5}   \,\,.       
\end{equation}   

\section{Discussion\label{disc}}

We start by comparing our results to those that can be found in the 
literature.
The result most frequently quoted is that of Donoghue, 
He, and Pakvasa~\cite{dhp} given in their Table~II, 
\begin{eqnarray}     
A\bigl(\Lambda^0_-\bigr) \,=\, -5 \times 10^{-5}  \,\,,  \hspace{3em}   
A\bigl(\Xi^-_-\bigr)  \,=\,  -7 \times 10^{-5}  \,\,.  
\label{dhpnum}
\end{eqnarray}
This result was computed using the matrix elements obtained 
by Donoghue, Golowich, Ponce, and Holstein~\cite{dghp}. 
Recast in the language of our previous sections,  Ref.~\cite{dghp}   
estimated  $\,{\rm Im}\, h_{D,F}^{}\,$  and  
$\,{\rm Im}\, \gamma_8^{}\,$  as the sum of direct and factorizable 
contributions in the same way we have done in this paper. The 
direct (nonfactorizable) contributions were calculated in the 
MIT bag model, and we agree with their results up to numerical 
inputs. The factorizable contributions in Ref.~\cite{dghp} are 
the ones they attribute to the quantity  ``${\cal O}_5^{(c)}$''.  
We disagree with the calculation of these factorizable terms in   
Ref.~\cite{dghp} in several important ways.

\begin{itemize}    
   
\item For the $S$-waves, we obtain a factorizable contribution to  
$h_F^{}$ approximately 4 times larger than that of Ref.~\cite{dghp}. 
This can be traced mainly to a difference in two factors. 
First, for the chiral condensate we use  
$\,\langle 0|\bar d  d|0\rangle = 
-f_\pi^2 B_0^{}\simeq -0.012\,\rm GeV^3,\,$  instead of 
$\,\langle 0|\bar d d|0\rangle = -0.007\,\rm GeV^3\,$  used in 
Ref.~\cite{dghp}. 
Second, we employ the value  $\,b_F^{}m_s^{}\simeq-95\,\rm MeV\,$   
in Eq.~(\ref{bDbF}), obtained from a first-order fit to 
the baryon-octet masses with Eq.~(\ref{Ls2}),  
whereas Ref.~\cite{dghp} calculate a baryon overlap in the MIT 
bag model that is equivalent to using   
$\,b_F^{}m_s^{}\simeq-43\,\rm MeV,\,$  
with  $\,m_s^{}=170\,\rm MeV.\,$   
     
\item A second difference in the $S$-wave phases 
(less important numerically) occurs because we use 
$\,h_D^{}\sim -0.3 h_F^{},\,$  as can be seen from 
Eqs.~(\ref{h[fac]}) and~(\ref{bDbFtree}),  
whereas the results of Ref.~\cite{dghp} used in Ref.~\cite{dhp} 
correspond to  $\,h_D^{}=0.\,$   
   
\item  Our most important difference occurs in the $P$-waves. 
Our factorization results from leading-order $\chi$PT 
calculations arise from the baryon poles. 
In contrast, the results of Ref.~\cite{dghp} for the baryon  
poles appear to include only the nonfactorizable contributions, 
and their $P$-waves are instead dominated by the kaon pole, 
as in Fig.~\ref{tree1}. 
This kaon pole is not included in our calculation 
because it occurs at next-to-leading order in $\chi$PT and, moreover, 
it is further suppressed by a factor of $m_\pi^2/m_K^2$ because 
the pion (and not the kaon) is on-shell. 
   
We have calculated this kaon-pole contribution (although we do  
not include it in our final results) and present it in 
the sixth column of our Table~\ref{spfac} under the heading 
``Im~$p_{\rm tree}^{(\gamma_8^{})}$''. 
It can be seen from this table that the kaon pole is 
indeed negligible compared to the baryon poles. 
Studying the calculation of Ref.~\cite{dghp}, we believe that 
their large result for the kaon pole is incorrect. 
The specific error arises in the evaluation of the 
kaon-pion weak transition in the bag model. 
We show some details in the last part of Appendix~\ref{bagmodel}.   
It is useful to cast this issue in the language 
adopted by the $\epsilon^\prime$ literature~\cite{buras},
\begin{equation}
\bigl\langle Q_6^{}\bigr\rangle_0^{}  \,=\, -4 \sqrt{\frac{3}{2}}\, 
\biggl[\frac{m_K^2}{m_s^{}(\mu) + m_d^{}(\mu)}\biggr]^2 
\bigl(f_K^{} - f_\pi^{}\bigr) B_6^{(1/2)}  \,\,,  
\label{b6buras}
\end{equation} 
where  
$\,\langle Q_6^{}\rangle_0\equiv\langle\pi\pi,I=0|Q_6^{}|K\rangle.\,$  
In our estimate, we use a $\gamma_8^{}$ corresponding to the 
value $\,B_6^{(1/2)}=1\,$ from factorization.  
For comparison, current lattice estimates are in the range 
$\,B_6^{(1/2)}=1\pm 1\,$~\cite{lattice},  whereas the calculation 
of Ref.~\cite{dghp} is equivalent to  $\,B_6^{(1/2)}\simeq 35.\,$
          
Despite this disagreement, the numerical value for the $P$-wave 
phases based on the results of Ref.~\cite{dghp} is similar to ours.  
This agreement is fortuitous and occurs because the factorizable 
contribution to the baryon poles is roughly equal to 35~times 
the kaon pole, as can be seen in Table~\ref{spfac}.   

\end{itemize}
   
In view of the above, the resulting numerical differences occur 
mostly in the $S$-wave phases, ours being larger than those 
found in Ref.~\cite{dhp}.   
This in turn impacts mainly the phase difference in the $\Lambda$ case, 
as  $\phi_{S,P}^\Lambda$  now tend to cancel each other. 
In contrast, the corresponding phase difference calculated 
using the results of Ref.~\cite{dghp} is much larger 
(by a factor of~5), being dominated by the $P$-wave phase.   
In the $\Xi$ case, the two weak phases have opposite signs, and 
so their difference is not suppressed, but instead it is now 
enhanced (by a factor of~3) with respect to that based 
on Ref.~\cite{dghp}.      
All these differences lead to the central values in Eq.~(\ref{A}), 
in comparison to the results of Ref.~\cite{dhp} 
in Eq.~(\ref{dhpnum}).     
An additional problem with the numbers in Eq.~(\ref{dhpnum}) is 
that they follow from outdated numerical input for the CKM matrix 
elements (and also from the use of the large old value  
$\,\delta_S^\Xi\sim-18^\circ\,$  for the $\Xi$  decay~\cite{nk}).

Next we turn our attention to the vacuum-saturation calculation 
of Ref.~\cite{hsv}.   
Our results in Tables~\ref{spbag} and~\ref{spfac} indicate that 
the factorization contribution is significantly larger than 
the direct contribution to the $S$- and $P$-wave phases. 
For this reason, we would expect our results to agree with those 
of  Ref.~\cite{hsv} in which the direct contributions are ignored. 
We find that we agree with the value of the $S$-wave phases up to 
numerical input, but that we disagree with the value of the 
$P$-wave phases in Ref.~\cite{hsv}. This disagreement is easy to  
understand. Our $P$-wave phases are dominated by the 
baryon-pole contribution, whereas in Ref.~\cite{hsv} only 
the kaon-pole contribution is included.  
The vacuum-saturation calculation of the kaon pole, corresponding to 
$B_6^{(1/2)}=1$, is a significant underestimate for the $P$-wave 
phases as seen in Table~\ref{spfac}, where the kaon pole corresponds 
to the column labeled ``Im~$p_{\rm tree}^{(\gamma_8^{})}$''.

To summarize then, the bag-model calculation of Ref.~\cite{dghp} 
significantly overestimates the contribution of the kaon pole to 
the $P$-waves and apparently misses the important factorization 
contribution of the baryon poles, although accidentally results in 
$P$-wave phases numerically similar to ours.   
Furthermore, it underestimates the $S$-waves and therefore 
yields an asymmetry dominated by the $P$-wave phase. 
The vacuum-saturation calculation of Ref.~\cite{hsv} misses 
the dominant baryon-pole contribution to the $P$-wave phases 
and results in an asymmetry dominated by the phase of the $S$-wave. 
In our complete calculation at leading order in $\chi$PT, the  
phases of the $S$- and $P$-waves are comparable, and in the 
$\Lambda$ case this leads to a smaller central value for the 
predicted asymmetry (the two phases tend to cancel).

It is difficult to place the calculation of Ref.~\cite{im} in our 
framework due to significant technical differences in the evaluation 
of loop integrals. Nevertheless, there is a rough correspondence 
between that calculation for the $S$-waves and the terms in 
Table~\ref{spfac} labeled  ``Im~$s_{\rm loop}^{(\gamma_8^{})}$''. 
In our final results, 
such terms appear in the quoted uncertainty because they are part 
of the subleading amplitudes that cannot be calculated completely 
at present.

In conclusion, we have presented a complete calculation of the 
weak phases in nonleptonic hyperon decay at leading order in 
heavy-baryon chiral perturbation theory. We have estimated the 
uncertainty in our calculation by computing the leading nonanalytic 
corrections. We have compared our results with those in the literature, 
pointing out several errors in previous calculations. To improve 
upon the results presented in this paper, it will be necessary to 
have a better understanding of the $P$-waves in nonleptonic hyperon decay.

\begin{acknowledgments}  
We would like to thank John F. Donoghue for useful discussions.  
The work of J.T. was supported in part by the U.S. Department of Energy
under contract DE-FG01-00ER45832 and by the Lightner-Sams Foundation.   
The work of G.V. was supported in part by the U.S. Department of Energy 
under contract DE-FG02-01ER41155.  
\end{acknowledgments}   
   
\appendix  
 
\section{Bag-model parameters\label{bagmodel}}  
  
In this appendix, we summarize the derivation of the formulas in 
Eq.~(\ref{hd,hf,hc}), which describe the nonfactorizable 
contributions to the weak parameters  $h_{D,F,C}^{}$,    
estimated in the MIT bag model.\footnote{An introductory treatment
of the bag model can be found in Ref.~\cite{dgh}} 
We also provide the numerical values of the parameters 
$a$ and $b$ in these formulas.   
Lastly, we evaluate the kaon-pion matrix element of the leading 
penguin operator in the bag model.

Assuming a valence-quark model of baryons, using the totally 
antisymmetric nature of their color wavefunctions and the 
relations~\cite{buras} 
\begin{eqnarray}   \label{4,9,10}    
Q_4^{}  \,=\,  -Q_1^{} + Q_2^{} + Q_3^{}   \,\,,   
\hspace{2em}    
Q_9^{}  \,=\,   
\mbox{$\frac{3}{2}$}\, Q_1^{} - \mbox{$\frac{1}{2}$}\, Q_3^{}   \,\,,   
\end{eqnarray}      
one finds for baryons  $B$  and  $B'$     
\begin{eqnarray}   \label{<B'|Q+Q'|B>}   
\begin{array}{c}   \displaystyle   
\bigl\langle B' \bigl| Q_1^{} \bigr| B \bigr\rangle 
\,=\,  - \bigl\langle B' \bigl| Q_2^{} \bigr| B \bigr\rangle 
\,=\,  \bigl\langle B' \bigl| Q_3^{} \bigr| B \bigr\rangle 
\,=\,  \bigl\langle B' \bigl| Q_9^{} \bigr| B \bigr\rangle 
\,=\,  - \bigl\langle B' \bigl| Q_{10}^{} \bigr| B \bigr\rangle   \,\,,    
\vspace{2ex} \\   \displaystyle    
\bigl\langle B' \bigl| Q_3^{}+Q_4^{} \bigr| B \bigr\rangle 
\,=\,  \bigl\langle B' \bigl| Q_5^{}+Q_6^{} \bigr| B \bigr\rangle 
\,=\,  \bigl\langle B' \bigl| Q_7^{}+Q_8^{} \bigr| B \bigr\rangle 
\,=\,  0   \,\,.      
\end{array}      
\end{eqnarray}      
Therefore, only  
$ \bigl\langle B' \bigl| Q_{1,5,7}^{} \bigr| B \bigr\rangle $  
need to be evaluated.   
For the parity-conserving parts of  $Q_{1,5,7}^{}$,  we  
derive the bag-model matrix elements\footnote{In keeping with 
Eq.~(\ref{<B'|H|B>}), we have excluded from these results the 27-plet  
components of $Q_{1,7}^{}$  and the  
$\bigl(8_{\rm L}^{},8_{\rm R}^{}\bigr)$ component of  $Q_7^{}$,  
the strong penguin operator  $Q_5^{}$  being purely   
$\bigl(8_{\rm L}^{},1_{\rm R}^{}\bigr)$.
Furthermore, in the $\Omega$-$\Xi^*$ matrix-elements we have taken 
into account the fact that the spinors for decuplet baryons in 
the chiral Lagrangian are spacelike~\cite{JenMan1},  
$\,\bar u_{\Xi^*}^{}\cdot u_\Omega^{}<0.\,$  
}   
\begin{eqnarray}   \label{<B'|Qi|B>}         
\begin{array}{c}   \displaystyle   
\bigl\langle n \bigl| Q_1^{} \bigr| \Lambda \bigr\rangle 
=  \bigl\langle n \bigl| Q_5^{} \bigr| \Lambda \bigr\rangle 
= -2 \bigl\langle n \bigl| Q_7^{} \bigr| \Lambda \bigr\rangle  
=  -\sqrt 6\, (a+b)   \,\,,   
\vspace{2ex} \\   \displaystyle    
\bigl\langle \Lambda \bigl| Q_1^{} \bigr| \Xi^0 \bigr\rangle 
=  2\sqrt 6\, (a+b)   \,\,,    
\hspace{3ex}  
\bigl\langle \Lambda \bigl| Q_5^{} \bigr| \Xi^0 \bigr\rangle 
= -2 \bigl\langle \Lambda \bigl| Q_7^{} \bigr| \Xi^0 \bigr\rangle 
=  \frac{8\sqrt 6\, b}{3}   \,\,,  
\vspace{2ex} \\   \displaystyle    
\bigl\langle \Xi^{*-} \bigl| Q_1^{} \bigr| \Omega^- \bigr\rangle 
=  0   \,\,,   
\hspace{3ex}    
\bigl\langle \Xi^{*-} \bigl| Q_5^{} \bigr| \Omega^- \bigr\rangle 
= -2 \bigl\langle \Xi^{*-} \bigl| Q_7^{} \bigr| \Omega^- \bigr\rangle 
= -4\sqrt 3 \Bigl( a-\mbox{$\frac{1}{3}$}\, b \Bigr)   \,\,,  
\end{array}      
\end{eqnarray}      
up to factors of  $\,\bar u_{B'}^{} u_B^{},\,$    
where  $a$  and  $b$  will be described shortly.  
From these results and Eq.~(\ref{<B'|H|B>}), we then obtain  
\begin{eqnarray}   \label{hd+3hf}
h_D^{} + 3 h_F^{}  \,=\,  
\frac{G_{\rm F}^{} \lambda}{\sqrt 2} \Bigl(   
C_1^{}-C_2^{}+C_3^{}-C_4^{}+C_9^{}-C_{10}^{} + C_5^{}-C_6^{}  
- \mbox{$\frac{1}{2}$} C_7^{}+\mbox{$\frac{1}{2}$} C_8^{} \Bigr) \, 
6 (-a-b)   \,\,,  
\end{eqnarray}      
\begin{eqnarray}   \label{hd-3hf}
h_D^{} - 3 h_F^{}  &=&    
\frac{G_{\rm F}^{} \lambda}{\sqrt 2} \Bigl[  
\bigl( C_1^{}-C_2^{}+C_3^{}-C_4^{}+C_9^{}-C_{10}^{} \bigr) \, 12(a+b)   
+ \Bigl( C_5^{}-C_6^{} - \mbox{$\frac{1}{2}$} C_7^{} 
        + \mbox{$\frac{1}{2}$} C_8^{} \Bigr) \, 16\, b 
\Bigr]    \,\,,  \nonumber \\  
\end{eqnarray}      
\begin{eqnarray} 
h_C^{}  \,=\,   
\frac{G_{\rm F}^{} \lambda}{\sqrt 2} 
\Bigl( C_5^{}-C_6^{} - \mbox{$\frac{1}{2}$} C_7^{} 
      + \mbox{$\frac{1}{2}$} C_8^{} \Bigr) \, (12\, a-4\, b)   \,\,.  
\end{eqnarray}      

The values of  $a$ and $b$  are found from the wavefunction overlap 
integrals  
\begin{eqnarray}   \label{a,b} 
a  \,=\,   
4\pi\int_0^R {\rm d}r\, r^2 \left( U^4(r)+L^4(r) \right)   \,\,,  
\hspace{3em}  
b  \,=\,  
8\pi\int_0^R {\rm d}r\, r^2\, U^2(r)\, L^2(r)   \,\,,   
\end{eqnarray}      
where  $R$  is the bag radius,  and   $U$  and  $L$  are the 
radial functions contained in the spatial wavefunctions 
\begin{eqnarray}    \label{psi,quark}     
\psi_q^{}(x)  \,\equiv\,  
\left( \begin{array}{c}   \displaystyle
{\rm i} U(r)\, \chi 
\vspace{2ex} \\   \displaystyle  
-L(r)\, \mbox{$\bm{\sigma}$}\!\cdot\!\hat{\mbox{$\bm{r}$}}\, \chi 
\end{array} \right)   \,\,,   
\hspace{2em}  
\psi_{\bar q}^{}(x)  \,\equiv\,  
\left( \begin{array}{c}   \displaystyle
-{\rm i} L(r)\, \mbox{$\bm{\sigma}$}\!\cdot\!\hat{\mbox{$\bm{r}$}}\, 
{\rm i}\sigma_y^{} \chi 
\vspace{2ex} \\   \displaystyle  
U(r)\, {\rm i}\sigma_y^{} \chi 
\end{array} \right)        
\end{eqnarray}      
of a quark $q$  and an antiquark  $\bar q$, respectively, 
with   $\chi$  being a two-component spinor  and      
$\sigma_i^{}$  the Pauli matrices.  
Explicitly,  $U(r)$  and  $L(r)$  are given in terms of spherical 
Bessel functions by~\cite{dgh}   
\begin{eqnarray}   
U(r)  \,=\,  
\frac{\cal N}{\sqrt{4\pi R^3}}\,\, j_0^{} \bigl( p r/R \bigr)   \,\,,   
\hspace{3em}   
L(r)  \,=\,   
\frac{\cal N}{\sqrt{4\pi R^3}}\,\, 
\epsilon\, j_1^{} \bigl( p r/R \bigr)   \,\,,   
\end{eqnarray}      
where  
\begin{eqnarray}   
{\cal N}  \,=\,  
\frac{p^2}{\sqrt{\left( 2\omega^2-2\omega+m R\right) \sin^2 p}}  \,\,, 
\hspace{2em}   
p  \,=\,  \sqrt{\omega^2-m^2 R^2}   \,\,,   
\hspace{2em}   
\epsilon  \,=\,  \sqrt{\frac{\omega-m R}{\omega+m R}}   \,\,,       
\end{eqnarray}      
with  $\omega$  being determined from  $\,\tan p=p/(1-\omega-m R)\,$   
and  $m$  the quark mass in the bag.  
Numerically, following Refs.~\cite{dghp,mitbagmodel}, we take  
$\,R=5.0\,{\rm GeV}^{-1}\,$  for  octet baryons  and  
$\,R=5.4\,{\rm GeV}^{-1}\,$  for  decuplet baryons.   
Since the weak parameters  $h_{D,F,C}^{}$  belong to a Lagrangian 
which respects SU(3) symmetry  
$\bigl[{\cal L}_{\rm w}^{}$  in Eq.~(\ref{weakcl})$\bigr]$,  
in writing  Eqs.~(\ref{<B'|Qi|B>})  and~(\ref{a,b})  
we have employed SU(3)-symmetric kinematics.\footnote{We note that in 
the SU(3)-symmetric limit the bag parameters above are related to  
the parameters  $A$  and  $B$  of Ref.~\cite{dghp} by   
$\,A=4\pi (a-b)R^3/{\cal N}^4\,$  and  $\,B=8\pi b R^3/{\cal N}^4.\,$}    
Specifically, we take  $\,m=0\,$  for all quark flavors.    
Thus, we find for octet baryons   
\begin{eqnarray}   \label{ob}
a  \,=\,  1.40\times 10^{-3}\, {\rm GeV}^3   \,\,,   \hspace{3em}   
b  \,=\,  0.64\times 10^{-3}\, {\rm GeV}^3   \,\,,    
\end{eqnarray}      
and for decuplet baryons   
\begin{eqnarray}   \label{db}
a  \,=\,  1.11\times 10^{-3}\, {\rm GeV}^3   \,\,,   \hspace{3em}   
b  \,=\,  0.51\times 10^{-3}\, {\rm GeV}^3   \,\,.   
\end{eqnarray}      

Finally, we evaluate the  $K$-to-$\pi$  transition in the bag model,  
which occurs in the kaon-pole result of Ref.~\cite{dghp},  
as discussed in our Sec.~\ref{disc}.
The matching of the dominant  $\,|\Delta I|=\frac{1}{2}\,$  
part of the weak Hamiltonian in Eq.~(\ref{weaksd}) to the weak chiral 
Lagrangian in Eq.~(\ref{weakcl}) involves in this case
\begin{eqnarray} 
\bigl\langle \pi^-(p) \bigr| \bigl( {\cal H}_{\rm w}^{} \bigr) _8^{}
\bigl| K^-(p) \bigr\rangle
\,=\,  -2 \gamma_8^{}\, p^2   \,\,.
\end{eqnarray}  
Concentrating on the $Q_6^{}$ contribution alone, we find 
the bag-model matrix element 
\begin{eqnarray}   \label{<pi|Q6|K>}     
\bigl\langle \pi^- \bigl| Q_6^{} \bigr| K^- \bigr\rangle 
\,=\,  -12(a + b)\, \sqrt{4 E_\pi^{} E_K^{}}   \,,         
\end{eqnarray}      
where the factor $\sqrt{4 E_\pi E_K}$  arises from  
the normalization of the bag states for the mesons~\cite{dgh,dghp}.  
From the preceding two equations, we obtain the $Q_6^{}$  
contribution   
\begin{eqnarray}   \label{g8bag}
\frac{\gamma_8^{} p^2}{\sqrt{4 E_\pi^{} E_K^{}}}  \,=\,    
\frac{G_{\rm F}^{} \lambda}{\sqrt 2}\, 
C_6^{}\, 6(a+b)   \,\,.  
\end{eqnarray}      
To determine the values of $a$ and $b$ in this equation, we use 
$\,R=3.3\,{\rm GeV}^{-1},\,$  after Ref.~\cite{dghp}, 
and again set  $\,m=0\,$  for all quark flavors. 
It follows that here  
\begin{eqnarray}   \label{om}
a  \,=\,  4.87\times 10^{-3}\, {\rm GeV}^3   \,\,,   \hspace{3em}   
b  \,=\,  2.23\times 10^{-3}\, {\rm GeV}^3   \,\,.   
\end{eqnarray}      
At this stage (in their equivalent calculation), Ref.~\cite{dghp}  
proceeds by setting  $\,p^2=m_\pi^2\,$  and  
$\,4 E_\pi^{} E_K^{}=2m_K^2.\,$ 
As a consequence, 
\begin{eqnarray} 
{\rm Im}\, \gamma_8^{}\, B_0^{}  \,=\,   
\frac{G_{\rm F}^{} \lambda}{\sqrt 2}\, {\rm Im}\, C_6^{}\,\,  
6 (a+b) \frac{\sqrt 2\, B_0^{}\, m_K^{}}{m_\pi^2}  
\,=\,  638\,\, y_6^{}     
\end{eqnarray}      
in units of   
$\,\sqrt 2\, f_\pi^{} G_{\rm F}^{} m_{\pi^+}^2\,\eta \lambda^5 A^2,\,$   
with the $B_0^{}$ value in Eq.~(\ref{bDbFtree}).  
Comparing this result with  Eq.~(\ref{Imh[fac]}) then indicates that  
the bag-model calculation of Ref.~\cite{dghp}  yields  
$\,B_6^{(1/2)}\sim 35,\,$   which is unacceptably large.

\section{Weak parameters in factorization\label{wpfac}} 
  
To derive the factorizable contributions to the imaginary part of 
the parameters  $h_{D,F,C}^{}$, we start from the observation that 
the quark-mass terms in the QCD Lagrangian can be written as  
\begin{eqnarray}   
{\cal L}_m^{}  \,=\, 
\frac{-1}{ 2 B_0^{}} \left( 
\bar q_{\rm L}^{}\, \chi\, q_{\rm R}^{}   
+ \bar q_{\rm R}^{}\, \chi^\dagger\, q_{\rm L}^{}
\right)   \,,  
\end{eqnarray}      
where  $\,q_{\rm L}^{}=\frac{1}{2}(1-\gamma_5)q\,$ 
and  $\,q_{\rm R}^{}=\frac{1}{2}(1+\gamma_5)q,\,$   
with  $\,q=(u\,\,\,d\,\,\,s)^{\rm T}.\,$    
It follows that  
\begin{eqnarray}   
\begin{array}{c}   \displaystyle   
-\bar q_{l\rm L}^{}\, q_{k\rm R}^{}  \,=\,  
2B_0^{}\, \frac{\delta {\cal L}_m^{}}{\delta \chi_{lk}^{}}   \,\,,  
\hspace{3em}   
-\bar q_{l\rm R}^{}\, q_{k\rm L}^{}  \,=\,  
2B_0^{}\, \frac{\delta {\cal L}_m^{}}{\delta \chi_{lk}^\dagger}   \,\,.    
\end{array}      
\end{eqnarray}      
Then, using  ${\cal L}_s^{(2,4)}$  in Eqs.~(\ref{Ls2})  and~(\ref{Ls4}),  
we have the correspondences  
\begin{eqnarray}   
-\bar q_{l\rm L}^{}\, q_{k\rm R}^{}  &\Longleftrightarrow&
b_D^{}\, 
\bigl( \xi^\dagger B^{} \bar B^{} \xi^\dagger 
      + \xi^\dagger \bar B^{} B^{} \xi^\dagger \bigr) _{kl}^{}
\,+\,  
b_F^{}\, 
\bigl( \xi^\dagger B^{} \bar B^{} \xi^\dagger   
      - \xi^\dagger \bar B^{} B^{} \xi^\dagger \bigr)_{kl}^{}  
\,+\,  
\sigma^{}\, \Sigma_{kl}^\dagger\, 
\bigl\langle \bar B^{} B^{} \bigr\rangle   
\nonumber \\ && 
+\,\,  
c\, \bigl( \bar T^\alpha \bigr) _{abc}^{}\,  
\xi_{cl}^\dagger \xi_{kd}^\dagger\, 
\bigl( T_\alpha^{} \bigr)_{dab}^{}      
\,-\,   
c_0^{}\, \Sigma_{kl}^\dagger\, \bar T^\alpha T_{\alpha}^{}  
\,\,+\,\,  
\mbox{$\frac{1}{2}$} f^2 B_0^{}\, \Sigma_{kl}^\dagger   
\nonumber \\ && 
+\,\,  
2 B_0^{} L_{5}^{} \left( \partial^\mu\Sigma^\dagger\, 
\partial_\mu^{}\Sigma\, \Sigma^\dagger \right)_{kl}^{}   
\,\,+\,\,  \cdots   \,\,,   
\end{eqnarray}      
\begin{eqnarray}   
-\bar q_{l\rm R}^{}\, q_{k\rm L}^{}  &\Longleftrightarrow&
b_D^{} \left( \xi B^{} \bar B^{} \xi 
             + \xi \bar B^{} B^{} \xi \right) _{kl}^{}
\,+\,  
b_F^{} \left( \xi B^{} \bar B^{} \xi   
             - \xi \bar B^{} B^{} \xi \right)_{kl}^{}  
\,+\,  
\sigma^{}\, \Sigma_{kl}^{}\, 
\bigl\langle \bar B^{} B^{} \bigr\rangle   
\nonumber \\ && 
+\,\,  
c\, \bigl( \bar T^\alpha \bigr) _{abc}^{}\,  
\xi_{cl}^{} \xi_{kd}^{}\, \bigl( T_\alpha^{} \bigr)_{dab}^{}      
\,-\,   
c_0^{}\, \Sigma_{kl}^{}\, \bar T^\alpha T_{\alpha}^{}  
\,\,+\,\,  
\mbox{$\frac{1}{2}$} f^2 B_0^{}\, \Sigma_{kl}^{}   
\nonumber \\ && 
+\,\,  
2 B_0^{} L_{5}^{} \left( \Sigma\, \partial^\mu\Sigma^\dagger\, 
\partial_\mu^{}\Sigma \right)_{kl}^{}   
\,\,+\,\,  \cdots   \,\,,   
\end{eqnarray}      
where the ellipses denote additional terms from  
${\cal L}_{\rm s}^{(4)}$ that do not affect our result. 
Consequently, for the penguin operator  
\begin{eqnarray}   
Q_6^{}  \,=\,  
-2 \sum_{q=u,d,s}\, \bar d (1+\gamma_5^{}) q\, 
\bar{q} (1-\gamma_5^{}) s   
\,=\,   
-8 \sum_{q=u,d,s}\, \bar d_{\rm L}^{} q_{\rm R}^{}\, 
\bar{q}_{\rm R}^{} s_{\rm L}^{}   \,\,,         
\end{eqnarray}      
we obtain  
\begin{eqnarray}   \label{Q6o}  
-\mbox{$\frac{1}{8}$} Q_6^{}  &\Longleftrightarrow&   
f^2 B_0^{} \left( 
b_D^{} \left\langle \bar B^{}\, 
\bigl\{ \xi^\dagger h \xi, B^{} \bigr\} \right\rangle  
\,+\,    
b_F^{} \left\langle \bar B^{}\, 
\bigl[ \xi^\dagger h \xi, B^{} \bigr] \right\rangle  
\right)     
\nonumber \\ &&  
+\,\,  f^2 B_0^{}\, c\, \bar T^\alpha\, \xi^\dagger h\xi\, T_\alpha^{}   
\,+\,  
2 f^2\, B_0^2 L_5^{} \left\langle h\, \partial_\mu^{}\Sigma\, 
\partial^\mu\Sigma^\dagger \right\rangle   
\,\,+\,\,  \cdots   \,\,,   
\end{eqnarray}      
where only the terms that correspond to leading-order chiral 
perturbation theory have been shown.  
Comparing this expression with the weak Lagrangian in 
Eq.~(\ref{weakcl}), we then infer that the contributions of 
a factorized $Q_6^{}$ to the weak parameters are 
\begin{eqnarray}   \label{h[fac]'}
\begin{array}{c}   \displaystyle   
h_D^{}  \,=\,  
\frac{G_{\rm F}^{}\, \lambda}{ \sqrt 2}\,\, 
8\, C_6^{}\, f^2 B_0^{}\, b_D^{}   \,\,,  
\hspace{3em}  
h_F^{}  \,=\,  
\frac{G_{\rm F}^{}\, \lambda}{\sqrt 2}\,\, 
8\, C_6^{}\, f^2 B_0^{}\, b_F^{}   \,\,,  
\vspace{2ex} \\   \displaystyle    
h_C^{}  \,=\,  
\frac{G_{\rm F}^{}\, \lambda}{\sqrt 2}\,\, 
8\, C_6^{}\, f^2 B_0^{}\, c   \,\,,  
\hspace{3em}  
\gamma_8^{}  \,=\,  
\frac{G_{\rm F}^{}\, \lambda}{\sqrt 2}\,\, 
16\, C_6^{}\, B_0^2\, L_5^{}   \,\,.  
\end{array}      
\end{eqnarray}      

\section{$\bm{CP}$-violating asymmetries in 
$\,\bm{\Sigma\to N\pi}\,$  decays\label{SigmaNpi}}  
    
The $S$-wave amplitudes in  $\,\Sigma\to N\pi\,$  can be expressed 
in terms of their components  $S_{2|\Delta I|,2I}^{}$, where the  $I$  
in the second subscript denotes the isospin of the $N\pi$ state.  
Thus we have\footnote{In the phase convention that 
we have adopted to write down these amplitudes, the isospin  
states  $\,|I,I_3^{}\rangle\,$  for the hadrons involved are  
$\,|\Sigma^+\rangle=-|1,1\rangle,\,$  $\,|\Sigma^-\rangle=|1,-1\rangle,\,$  
$\,|p\rangle=|1/2,1/2\rangle,\,$  $\,|n\rangle=|1/2,-1/2\rangle,\,$  
$\,|\pi^+\rangle=-|1,1\rangle,\,$  $\,|\pi^0\rangle= |1,0\rangle,\,$  
and   $\,|\pi^-\rangle= |1,-1\rangle,\,$  
which are consistent with the structure of the  $\varphi$  
and  $B_v^{}$  matrices in the chiral Lagrangian.}  
\begin{eqnarray}   \label{sSigma}
\begin{array}{c}   \displaystyle
s_{\Sigma^+\to n\pi^+}^{}  \,=\,  
\mbox{$\frac{1}{3} $} \Bigl( 
2 S_{11}^{}\, {\rm e}^{{\rm i}\phi_{11}^S} 
+ S_{31}^{}\, {\rm e}^{{\rm i}\phi_{31}^S} \Bigr) \, 
{\rm e}^{{\rm i}\delta_1^S}  
+   
\mbox{$\frac{1}{3}$} \left( S_{13}^{}\, {\rm e}^{{\rm i}\phi_{13}^S}
- 2\sqrt{\mbox{$\frac{2}{5}$}}\, S_{33}^{}\, {\rm e}^{{\rm i}\phi_{33}^S}   
\right) {\rm e}^{{\rm i}\delta_3^S}   \,\,,   
\vspace{2ex} \\   \displaystyle
s_{\Sigma^+\to p\pi^0}^{}  \,=\,
\mbox{$\frac{1}{3\sqrt 2}$} \Bigl( 
2 S_{11}^{}\, {\rm e}^{{\rm i}\phi_{11}^S} 
+ S_{31}^{}\, {\rm e}^{{\rm i}\phi_{31}^S} \Bigr) \, 
{\rm e}^{{\rm i}\delta_1^S}    
- \mbox{$\frac{\sqrt 2}{3}$} \left( 
S_{13}^{}\, {\rm e}^{{\rm i}\phi_{13}^S}
- 2\sqrt{\mbox{$\frac{2}{5}$}}\,  S_{33}^{}\, 
 {\rm e}^{{\rm i}\phi_{33}^S} 
\right) {\rm e}^{{\rm i}\delta_3^S}   \,\,,  
\vspace{2ex} \\   \displaystyle
s_{\Sigma^-\to n\pi^-}^{}  \,=\,
\left( S_{13}^{}\, {\rm e}^{{\rm i}\phi_{13}^S}  
+ \sqrt{\mbox{$\frac{2}{5}$}}\, S_{33}^{}\, {\rm e}^{{\rm i}\phi_{33}^S} 
\right) {\rm e}^{{\rm i}\delta_3^S}   \,\,,  
\end{array}
\end{eqnarray}
where  $\delta_{2I}^{S}$  and  $\,\phi_{2|\Delta I|,2I}^{S}\,$   
are  the strong $N\pi$-scattering and weak $CP$-violating phases, 
respectively,   
and  $\,|\Delta I|=\frac{5}{2}\,$  components have been ignored.  
The $P$-wave amplitudes can be similarly expressed.      
For each of these  decays,  one can construct 
the counterpart of the $CP$-violating asymmetries 
$A_{\Lambda,\Xi}^{}$  using~\cite{dhp}      
\begin{eqnarray} 
A  \,=\,  
\frac{\Gamma\,\alpha+\bar{\Gamma}\,\bar{\alpha}}
{\Gamma\,\alpha-\bar{\Gamma}\,\bar{\alpha}}   \,\,.    
\end{eqnarray}      
One then has  
\begin{eqnarray}    
A\bigl(\Sigma_+^+\bigr)  &\equiv&  A_{\Sigma^+\to n\pi^+}^{}  
\nonumber \\ &=& 
- \Biggl[  
\sin\bigl(\delta_1^P-\delta_1^S\bigr)\,   
\sin\bigl(\phi_{1}^P-\phi_{1}^S\bigr)      
+ \frac{S_{3}^{}}{2 S_{1}^{}}\, \sin\bigl(\delta_1^P-\delta_3^S\bigr)\,   
\sin\bigl(\phi_{1}^P-\phi_{3}^S\bigr)      
\nonumber \\ && \hspace{3ex}  
+\, \frac{P_{3}^{}}{2 P_{1}^{}}\, 
   \sin\bigl(\delta_3^P-\delta_1^S\bigr)\,   
\sin\bigl(\phi_{3}^P-\phi_{1}^S\bigr)      
+ \frac{S_{3}^{} P_{3}^{}}{4 S_{1}^{} P_{1}^{}}\,  
\sin\bigl(\delta_3^P-\delta_3^S\bigr)\,   
\sin\bigl(\phi_{3}^P-\phi_{3}^S\bigr)      
\Biggr] \Big/ 
\nonumber \\ && 
\Biggl[  
\cos\bigl(\delta_1^P-\delta_1^S\bigr)   
+ \frac{S_{3}^{}}{2 S_{1}^{}}\, \cos\bigl(\delta_1^P-\delta_3^S\bigr)   
+ \frac{P_{3}^{}}{2 P_{1}^{}}\, \cos\bigl(\delta_3^P-\delta_1^S\bigr)   
+ \frac{S_{3}^{} P_{3}^{}}{4 S_{1}^{} P_{1}^{}}\,    
\cos\bigl(\delta_3^P-\delta_3^S\bigr)     
\Biggr]   \,\,,  \hspace{2em}  
\end{eqnarray}      
\begin{eqnarray}    
A\bigl(\Sigma_0^+\bigr)  &\equiv&  A_{\Sigma^+\to p\pi^0}^{}  
\nonumber \\ &=&    
- \Biggl[  
\sin\bigl(\delta_1^P-\delta_1^S\bigr)\,   
\sin\bigl(\phi_{1}^P-\phi_{1}^S\bigr)      
- \frac{S_{3}^{}}{S_{1}^{}}\, \sin\bigl(\delta_1^P-\delta_3^S\bigr)\,   
\sin\bigl(\phi_{1}^P-\phi_{3}^S\bigr)      
\nonumber \\ && \hspace{3ex}  
-\, \frac{P_{3}^{}}{P_{1}^{}}\, 
   \sin\bigl(\delta_3^P-\delta_1^S\bigr)\,   
\sin\bigl(\phi_{3}^P-\phi_{1}^S\bigr)      
+ \frac{S_{3}^{} P_{3}^{}}{S_{1}^{} P_{1}^{}}\,  
\sin\bigl(\delta_3^P-\delta_3^S\bigr)\,   
\sin\bigl(\phi_{3}^P-\phi_{3}^S\bigr)      
\Biggr] \Big/ 
\nonumber \\ &&    
\Biggl[  
\cos\bigl(\delta_1^P-\delta_1^S\bigr)   
- \frac{S_{3}^{}}{S_{1}^{}}\, \cos\bigl(\delta_1^P-\delta_3^S\bigr)   
- \frac{P_{3}^{}}{P_{1}^{}}\, \cos\bigl(\delta_3^P-\delta_1^S\bigr)   
+ \frac{S_{3}^{} P_{3}^{}}{S_{1}^{} P_{1}^{}}\, 
\cos\bigl(\delta_3^P-\delta_3^S\bigr)     
\Biggr]   \,\,,  \hspace{2em}  
\end{eqnarray}
\begin{eqnarray}     
A\bigl(\Sigma_-^-\bigr)  \,\equiv\,  A_{\Sigma^-\to n\pi^-}^{}  
\,=\,  
- \tan\bigl(\delta_3^P-\delta_3^S\bigr)\, 
\frac{ \sin\bigl(\phi_{13}^P-\phi_{13}^S\bigr)      
+ \displaystyle\sqrt{\frac{2}{5}}\, \frac{S_{33}^{}}{S_{13}^{}}\, 
 \sin \phi_{13}^P      
- \displaystyle\sqrt{\frac{2}{5}}\, \frac{P_{33}^{}}{P_{13}^{}}\, 
 \sin \phi_{13}^S      
}{
1 + \displaystyle\sqrt{\frac{2}{5}}\, \frac{S_{33}^{}}{S_{13}^{}}  
+ \displaystyle\sqrt{\frac{2}{5}}\, \frac{P_{33}^{}}{P_{13}^{}}  
+ \displaystyle\frac{2\, S_{33}^{}P_{33}^{}}{5\, S_{13}^{}P_{13}^{}}  
}   \,\,,   \hspace{2em} 
\end{eqnarray}   
where    
\begin{eqnarray}   \label{13}
\begin{array}{c}   \displaystyle
S_1^{}  \,\equiv\,  S_{11}^{} + \mbox{$\frac{1}{2}$}\, S_{31}^{}   \,\,,   
\hspace{2em}  
\phi_1^{S}  \,\equiv\,  \frac{S_{11}^{}\, \phi_{11}^S}{S_1^{}}   \,\,,  
\hspace{3em}  
S_3^{}  \,\equiv\,  
S_{13}^{} - 2\sqrt{\mbox{$\frac{2}{5}$}}\, S_{33}^{}   \,\,,  
\hspace{2em}  
\phi_3^{S}  \,\equiv\,  
\frac{S_{13}^{}\, \phi_{13}^S}{S_3^{}}   \,\,,  
\end{array}    
\end{eqnarray}
the $P$-wave counterparts being similarly defined, and the weak  
$\,|\Delta I|=\frac{3}{2}\,$  phases have been neglected.

To estimate the weak phases,  we follow the prescription proposed 
earlier, obtaining the real part of the amplitudes from the values 
extracted from experiment under the assumption of no $CP$-violation
and calculating the imaginary part from the leading-order amplitudes 
in Eq.~(\ref{cptamp}) with the values of 
$\,{\rm Im}\,h_{D,F}^{}\,$  provided in Section~\ref{ect}.   
To find the real part, ignoring the strong and weak phases,  
we first derive from Eq.~(\ref{sSigma})    
\begin{eqnarray}   \label{s1}
\begin{array}{c}   \displaystyle
S_{1}^{}  \,=\,     
s_{\Sigma^+\to n\pi^+}^{} 
+ \mbox{$\frac{1}{\sqrt 2}$}\, s_{\Sigma^+\to p\pi^0}^{}   \,\,,  
\hspace{3em}  
3\, S_{13}^{}  \,=\,  
s_{\Sigma^+\to n\pi^+}^{} - \sqrt 2\, s_{\Sigma^+\to p\pi^0}^{}
+ 2\, s_{\Sigma^-\to n\pi^-}^{}   \,\,,
\vspace{2ex} \\   \displaystyle
S_{33}^{}  \,=\,  
-\sqrt{\mbox{$\frac{5}{18}$}}\, \Bigl( 
s_{\Sigma^+\to n\pi^+}^{} - \sqrt 2\, s_{\Sigma^+\to p\pi^0}^{}
- s_{\Sigma^-\to n\pi^-}^{} \Bigr)   \,\,,
\end{array}
\end{eqnarray}   
and analogous expressions for the $P$-waves.   
From the experimental values in Table~\ref{spx}, we then extract, 
in units of~$G_{\rm F}^{}m_{\pi^+}^2$,      
\begin{eqnarray}   \label{SPSigma}
\begin{array}{c}   \displaystyle      
S_{1}^{}  \,=\,  -0.95 \pm 0.04   \,\,,    
\hspace{2em}   
S_{13}^{}   \,=\,  1.95 \pm 0.02   \,\,,     
\hspace{2em}   
S_{33}^{}   \,=\,  -0.11 \pm 0.04   \,\,,     
\vspace{2ex} \\   \displaystyle
P_{1}^{}  \,=\,  2.64 \pm 0.04   \,\,,    
\hspace{2em}   
P_{13}^{}   \,=\,  0.01 \pm 0.03   \,\,,  
\hspace{2em}   
P_{33}^{}   \,=\,  -0.11 \pm 0.05   \,\,,     
\end{array}      
\end{eqnarray}      
The imaginary part of the amplitudes are obtained using 
Eqs.~(\ref{cptamp}), (\ref{Imh[bag]}), and~(\ref{Imh[fac]}), 
as well as the isospin relation
$\,\sqrt 2\, a_{\Sigma^+\to p\pi^0}^{}=
a_{\Sigma^+\to n\pi^+}^{} - a_{\Sigma^-\to n\pi^-}^{}\,$  
for $\,|\Delta I|=\frac{1}{2}\,$  dominance.     
Thus, we have  in units of  $\eta \lambda^5 A^2$ 
\begin{eqnarray}   \label{imSP}
\begin{array}{c}   \displaystyle      
\frac{{\rm Im}\, S_{1}^{}}{S_{1}^{\rm expt}}  \,=\,  
(-0.04+1.02) + (-0.03+1.30)   \,\,,    
\hspace{2em}   
\frac{{\rm Im}\, S_{13}^{}}{S_{13}^{\rm expt}}  \,=\,  
(-0.04+0.99) + (-0.03+1.26)  \,\,,     
\vspace{2ex} \\   \displaystyle
\frac{{\rm Im}\, P_{1}^{}}{P_{1}^{\rm expt}}  \,=\,  
(0.02+0.09)  + (-0.07+0.31)   \,\,,    
\hspace{2em}   
\frac{{\rm Im}\, P_{13}^{}}{P_{13}^{\rm expt}}   \,=\,   
(7-41) + (-15-59)   \,\,,     
\end{array}      
\end{eqnarray}      
where  the numerators on the left-hand sides are 
the central values in Eq.~(\ref{SPSigma}), and we have written 
each result as  $\,$(tree)$+$(loop),$\,$  with the two numbers  
within each pair of brackets being bag-model and factorization   
contributions, respectively. 
In Table~\ref{phases_Sigma}, we collect the weak phases resulting 
from these ratios.

We also show in Table~\ref{phases_Sigma} another error estimate, 
$\delta\phi$, obtained from using the leading-order amplitudes 
and allowing the parameters to vary between their tree-level and 
one-loop values, as discussed in Sec.~\ref{results}.      
In making this estimate, we again employ only the factorization 
contributions [for the $P$-waves, we use the $\Sigma$ amplitudes in 
Eq.~(\ref{Pfac})], which are are much larger than the bag-model 
ones, as seen in Eq.~(\ref{imSP}).  

\begin{table}[ht]   
\caption{\label{phases_Sigma}%
Weak $S$- and  $P$-wave phases in  $\,\Sigma\to N\pi\,$  decays  
from  $Q_6^{}$ contribution alone,  in units of  $\eta \lambda^5 A^2$.    
}  
\centering   \footnotesize
\vskip 0.5\baselineskip    
\begin{tabular}{ccc|ccc||ccc|ccc}
\hline \hline    
$\vphantom{\bigg|}$$\,\,\phi_{1}^{S,\rm tree}\,\,$  &  
$\,\,\phi_{1}^{S,\rm loop}\,\,$  &
$\,\,\delta\phi_{1}^{S,\rm tree}\,\,$  &   
$\,\,\phi_{13}^{S,\rm tree}\,\,$  &  $\,\,\phi_{13}^{S,\rm loop}\,\,$  &
$\,\,\delta\phi_{13}^{S,\rm tree}\,\,$  &   
$\,\,\phi_{11}^{P,\rm tree}\,\,$  &  $\,\,\phi_{11}^{P,\rm loop}\,\,$  &
$\,\,\delta\phi_{11}^{P,\rm tree}\,\,$  &   
$\,\,\phi_{13}^{P,\rm tree}\,\,$  &  $\,\,\phi_{13}^{P,\rm loop}\,\,$  &
$\,\,\delta\phi_{13}^{P,\rm tree}\,\,$     
\\ \hline && && && && && & \vspace{-3ex} \\     
0.98 & 1.27  &  $-$1.65  &  0.95 & 1.23  &  $-$1.61  &  
0.11 & 0.24  &  $-$0.05  &  $-$34 & $-$74  &  24   
\\   \hline \hline  
\end{tabular}   
\vspace{2ex}  
\end{table}   

We may, therefore, conclude that the uncertainties of the weak 
phases are all of order~200$\%$.   
This is reflected in our prediction of the phases, which are 
collected in Table~\ref{phases_Sigma'}.   
The corresponding strong phases have been measured~\cite{roper}    
and their values have also been included in this table.

\begin{table}[ht]   
\caption{\label{phases_Sigma'}%
Predicted weak phases, in units of  $\eta \lambda^5 A^2$, 
and measured strong phases. 
}  
\centering   \footnotesize
\vskip 0.5\baselineskip    
\begin{tabular}    
{@{\hspace{1ex}}c@{\hspace{3ex}}c@{\hspace{3ex}}c@{\hspace{3ex}}
c@{\hspace{2ex}}||@{\hspace{2ex}}c@{\hspace{3ex}}c@{\hspace{3ex}}
c@{\hspace{3ex}}c@{\hspace{1ex}}}   
\hline \hline    
$\vphantom{\bigg|}$$\,\,\phi_{1}^{S}\,\,$  &  
$\,\,\phi_{13}^{S}\,\,$  &  $\,\,\phi_{1}^{P}\,\,$  &  
$\,\,\phi_{13}^{P}\,\,$  &  
$\,\,\delta_1^{S}\,\,$  &  $\,\,\delta_3^{S}\,\,$  &  
$\,\,\delta_1^{P}\,\,$  &  $\,\,\delta_3^{P}\,\,$  
\\ \hline && & \vspace{-3ex} \\     
$1.0 \pm 2.0$  &  $1.0 \pm 2.0$  &  $0.1 \pm 0.2$  &  $-40 \pm 80$  &  
$9.4^\circ\pm 1.0^\circ$  &  $-10.1^\circ\pm 1.0^\circ$  &  
$-1.8^\circ\pm 1.0^\circ$  &  $-3.5^\circ\pm 1.0^\circ$  
\\   \hline \hline  
\end{tabular}   
\vspace{2ex}  
\end{table}

From the central values of the isospin amplitudes and the phases 
in  Eq.~(\ref{SPSigma}) and  Table~\ref{phases_Sigma'}, 
respectively, we obtain  
\begin{eqnarray}   \label{A_sigma}
\begin{array}{c}   \displaystyle      
A\bigl(\Sigma_+^+\bigr)  \,=\,  3.9\times 10^{-4}    \,\,,  
\hspace{3em}  
A\bigl(\Sigma_0^+\bigr)  \,=\,  3.6\times 10^{-6}   \,\,,  
\hspace{3em}  
A\bigl(\Sigma_-^-\bigr)  \,=\,  -8.3\times 10^{-5}    \,\,,     
\end{array}   
\end{eqnarray}   
where we have used  $\,A^2\lambda^5\eta=1.26\times 10^{-4}\,$  
as before.   
In this case our estimate is a very rough one, as its uncertainty 
is larger than those for the other hyperons.   
This is due to the (apparently accidental) smallness of  
$P_{13}^{}$ and its large experimental error, indicated  
in Eq.~(\ref{SPSigma}), as well as  to the already sizable 
uncertainties quoted in Table~\ref{phases_Sigma'}.   
In order to have a more quantitative estimate of the uncertainties, 
these modes will have to be revisited when better measurements of 
the amplitudes become available.

\end{document}